\documentclass[a4paper,11pt]{article}
\pdfoutput=1 % if your are submitting a pdflatex (i.e. if you have
\usepackage{jheppub,bm} % for details on the use of the package, please
\usepackage[T1]{fontenc} % if needed
\usepackage[utf8]{inputenc}

\usepackage{breqn} % line break for equation
\usepackage{graphicx} % Required for including images
\usepackage{subcaption} % Required for creating subfigures

\def\bI{\boldsymbol{I}}

\newcommand{\eps}{\epsilon}

\def\as{\alpha_s}

\def\beq{\begin{equation}}
\def\eeq{\end{equation}}
\def\beqn{\begin{eqnarray}}
\def\eeqn{\end{eqnarray}}

\def\taucm{\tau^{\rm{cut}}_1}

\def\spa#1.#2{\langle #1 #2\rangle}
\def\spb#1.#2{[ #1 #2]}
\def\spab#1.#2.#3{\langle #1 |#2| #3] }
\allowdisplaybreaks[4]

\begin{document}

\title{Pseudoscalar Higgs plus jet production at Next-to-Next-to-Leading Order in QCD}

\author{Youngjin Kim}
\author{and Ciaran Williams}

\affiliation{Department of Physics,\\ University at Buffalo, The State University of New York, Buffalo
14260, USA}

\emailAdd{ykim56@buffalo.edu}
\emailAdd{ciaranwi@buffalo.edu}

\newcommand{\zero}{{(0)}}
\newcommand{\one}{{(1)}}
\newcommand{\two}{{(2)}}
\newcommand{\ztwo}{\zeta_2}
\newcommand{\zthree}{\zeta_3}
\newcommand{\cf}{C_F}
\newcommand{\ca}{C_A}
\newcommand{\nf}{n_f}
\newcommand{\cfs}{C_F^2}
\newcommand{\Tcm}{\tau_{cm}}
\newcommand{\TAij}{\mathbf{T}^a_{ij}}

\abstract{
We present a calculation of pseudoscalar Higgs production in association with a jet at Next-to-Next-to Leading Order (NNLO) accuracy in QCD. We work in an effective field theory in which  $m_t \rightarrow \infty$ resulting in effective operators which couple the pseudoscalar to gluons and (massless) quarks. We have calculated all of the relevant amplitudes for the two-loop, one-loop and tree-level contributions. As a cross-check of our calculation we have re-calculated all of the scalar Higgs plus parton amplitudes and perform a detailed comparison to the literature. In order to regulate the infra-red singularities present at this order we employ the $N-$jettiness slicing method. In addition to a detailed validation of our calculation at this order we investigate LHC phenomenology for a selection of pseudoscalar Higgs masses. Our results are implemented into the parton-level Monte Carlo code MCFM. 
}
\maketitle

\flushbottom

\section{Introduction}

The continued study of the Higgs boson, discovered over a decade ago~\cite{ATLAS:2012yve,CMS:2012qbp}, remains one of the top priorities for high energy physics. Over the next twenty years the full data set of the HL-LHC will allow for unprecedented glimpse at the Higgs boson's coupling to matter, other bosons and ultimately itself. In all instances the hope is that deviations from the predictions of the Standard Model (SM) may be observed suggestive of new interactions which may elucidate the origins of Electroweak symmetry breaking.  In order to achieve this aim, precision predictions for the Higgs boson will be essential.
For the inclusive Higgs situation calculations have achieved a high level of precision with N3LO predictions completed~\cite{Anastasiou:2015vya,Mistlberger:2018etf,Chen:2021isd}.
At the LHC an observable of great interest is the transverse momentum of the recoiling Higgs boson. The calculation of the production of a Higgs boson at non-zero $p_T$ has been a successful area of research over the last ten years, with several groups pushing the accuracy to NNLO in QCD in an effective field theory in which the top quark mass goes to infinity~\cite{Boughezal:2013uia,Chen:2014gva,Boughezal:2015dra,Boughezal:2015aha,Caola:2015wna,Chen:2016zka,Campbell:2019gmd}. 
The Higgs effective theory with $m_t \to \infty$ provides, for the most part, an excellent description of the full theory. However, results from the EFT breakdown in certain regions of phase space (typically high-$p_T$) where a sensitivity to the top mass returns. The exact calculation the Higgs production including the top-quark mass dependence is known up to NLO in QCD~\cite{Spira:1995rr}.
The EFT prediction can be improved by including sub-leading top-quark mass corrections systematically as an expansion in the inverse top-quark mass~\cite{Pak:2009dg,Harlander:2009mq}.

Although the SM is a beautiful quantum field theory, and the discovery of the Higgs provided the last missing piece, there are well motivated reasons to believe it is not the fundamental theory of nature. Leading concerns include, the lack of a dark matter candidate, the ability to explain neutrino masses, or the amount of CP violation observed in the universe. For these reasons and more many theories Beyond the Standard Model (BSM) have been constructed to address and solve these problems. A common feature of a vast majority of BSM models includes an extended Higgs sector, for instance, any type of two Higgs doublet models or the Minimal Supersymmetric Standard Model~\cite{Fayet:1974pd,Dimopoulos:1981zb,Sakai:1981gr,Inoue:1982pi,Inoue:1983pp,Inoue:1982ej} introduce an extra scalar doublet.
In BSM theories which introduce a second Higgs doublet, after electroweak symmetry breaking, there are five physical scalar bosons which include two CP even Higgs states $h, H$, a CP odd pseudoscalar $A$ and two charged Higgs bosons $H^{\pm}$. Given the SM-like nature of (one of the) CP even Higgs bosons, extensions of the SM are naturally constrained such that the spectrum of states is in an alignment limit with the SM. 

 Multiple new scalar bosons is thus a simple, and somewhat model independent, prediction of extended Higgs sectors, and this motivated many dedicated experimental searches and theoretical work. In this paper we are primarily interested in the psuedoscalar $A$ boson (focusing on the situation where the
 ratio between the vacuum expectation of the Higgs doublets, $\tan{\beta}$ is small).  Given the mathematical similarities between the CP even and CP odd Higgs bosons, precision predictions for psuedoscalar bosons tend to track their CP-even Higgs counterparts. For the effective field theory induced by taking $m_t \rightarrow \infty$ scalar Higgs production at  next to leading order~\cite{Dawson:1990zj,Djouadi:1991tka} was calculated in the early 1990s, with the corresponding studies for the pseudoscalar Higgs production case following shortly after~\cite{Kauffman:1993nv}\footnote{The exact top mass dependence for pseudo scalar Higgs is known up to NLO~\cite{Spira:1993bb}.}. 
Later, in the early 2000s, the Next-to-Next-to-leading order (NNLO) predictions were completed~\cite{Harlander:2002wh,Anastasiou:2002yz,Ravindran:2003um,Anastasiou:2005qj}, immediately followed by the pseudoscalar Higgs case~\cite{Harlander:2002vv,Anastasiou:2002wq}. Inclusive Higgs production has recently been computed at N3LO~\cite{Anastasiou:2015vya,Mistlberger:2018etf,Chen:2021isd}, and some progress has been made towards the full N3LO predictions for pseudoscalar production~\cite{Ahmed:2015qda,Ahmed:2016otz}.
The corresponding studies of the (pseudo)scalar Higgs boson produced with an associated jet show a similar pattern. NLO predictions can be found in Refs.~\cite{deFlorian:1999zd,Ravindran:2002dc,Field:2002pb,Bernreuther:2010uw}. 
However, after a series of recent calculations of the production of the scalar Higgs plus jet at Next-to-Next-to leading order~\cite{Boughezal:2013uia,Chen:2014gva,Boughezal:2015dra,Boughezal:2015aha,
Caola:2015wna,Chen:2016zka,Campbell:2019gmd}, no such calculation for pseudoscalar Higgs bosons is available. Addressing this gap is the primary motivation of our paper. 

The NNLO  calculations of Higgs plus jet have relied on a series of breakthroughs over the last decade.
For a long time NNLO calculations at hadron colliders involving final state jets were beyond reach due to the technical complexities of infra-red (IR) limits. However, over the last ten years dramatic improvements have been made in this regard. Two popular techniques for dealing with IR singularities include N-jettiness slicing~\cite{Gaunt:2015pea,Boughezal:2015dva} and antenna subtraction~\cite{Gehrmann-DeRidder:2005btv,Daleo:2006xa,Daleo:2009yj,Gehrmann:2011wi,Boughezal:2010mc,Gehrmann-DeRidder:2012too,Currie:2013vh} and these methods have led to dramatic improvements in the number of NNLO predictions available.  In our calculation, we have taken advantage of the well-established MCFM~\cite{Boughezal:2016wmq,Campbell:2019dru} framework (which uses $N$-jettiness slicing) that has been successfully deployed for many calculations at NNLO, the most relevant for this work being the calculation of $H+j$ at NNLO~\cite{Campbell:2019gmd} and bottom induced Higgs plus jet production~\cite{Mondini:2021nck}.

Our paper proceeds as follows. In Section \ref{sec:overview} we give a general overview of the calculation, while a detailed discussion of our two-loop hard function calculation is presented in Section \ref{sec:hardfunction}, and the calculation of $A+2j$ at NLO is discussed in  Section~\ref{sec:abovecut}. 
We discuss the results of our Monte Carlo implementation of MCFM in Section \ref{sec:results} including the phenomenology for this process at the 13 TeV LHC. 
Then, lastly, we draw our conclusions in Section~\ref{sec:conclusion}. A series of useful formulae, and a detailed comparison with the $H+j$ calculation, as well as a description of ancillary files make up a series of Appendices.

%%%%%%%%%%%%%%%%%%%%%%%%%%Overview of the calculation%%%%%%%%%%%%%%%%%%%%%%%%%%%5
\section{Overview of the calculation}

\label{sec:overview}

\begin{figure}
\begin{center}
\centering % Centers the figure
% First image with a specific scale
\begin{subfigure}[b]{0.4\textwidth}
\centering % Center the image in the subfigure
\includegraphics[scale=0.04]{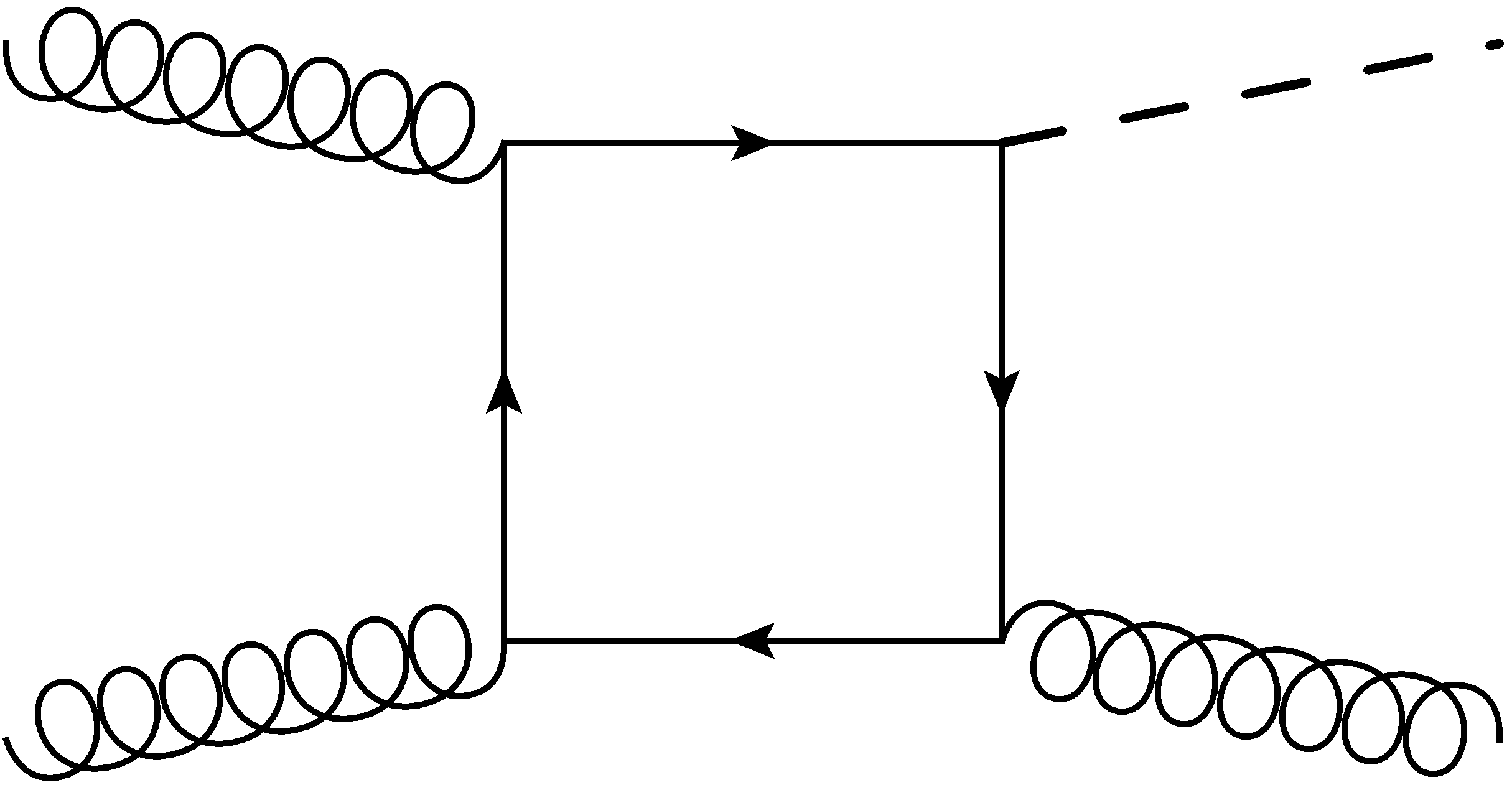} % Adjust scale as needed
\caption{$f = ggg$}
\label{fig:CJ_ggg}
\end{subfigure}
%\hfill % Adds horizontal space between the figures
% Second image with the same scale
\begin{subfigure}[b]{0.4\textwidth}
\centering % Center the image in the subfigure
\includegraphics[scale=0.04]{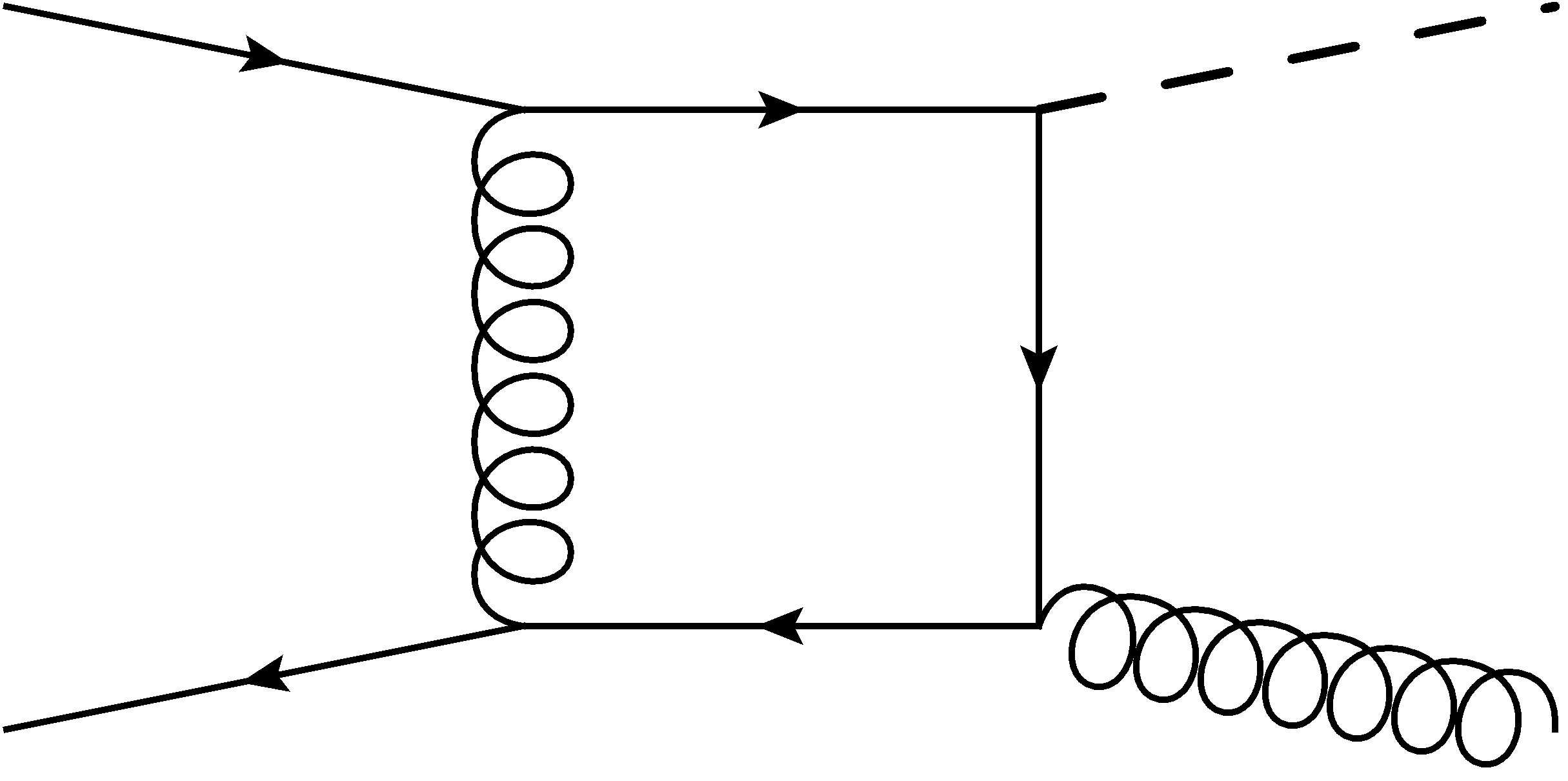} % Use the same scale
\caption{$f = q\bar{q}g$}
\label{fig:CJ_qqg}
\end{subfigure}
\hfill % Adds horizontal space between the figures
\caption{Representative $O_J$ Feynman diagrams which contribute at NNLO.} 
\label{feynman_diagramsOJ}
\end{center}
\end{figure}

In this paper we consider the production of a pseudoscalar Higgs boson $(A)$ and an additional jet 
up to NNLO in QCD.
\begin{figure}
\begin{center}
\centering % Centers the figure
% First image with a specific scale
\begin{subfigure}[b]{0.3\textwidth}
\centering % Center the image in the subfigure
\includegraphics[scale=0.04]{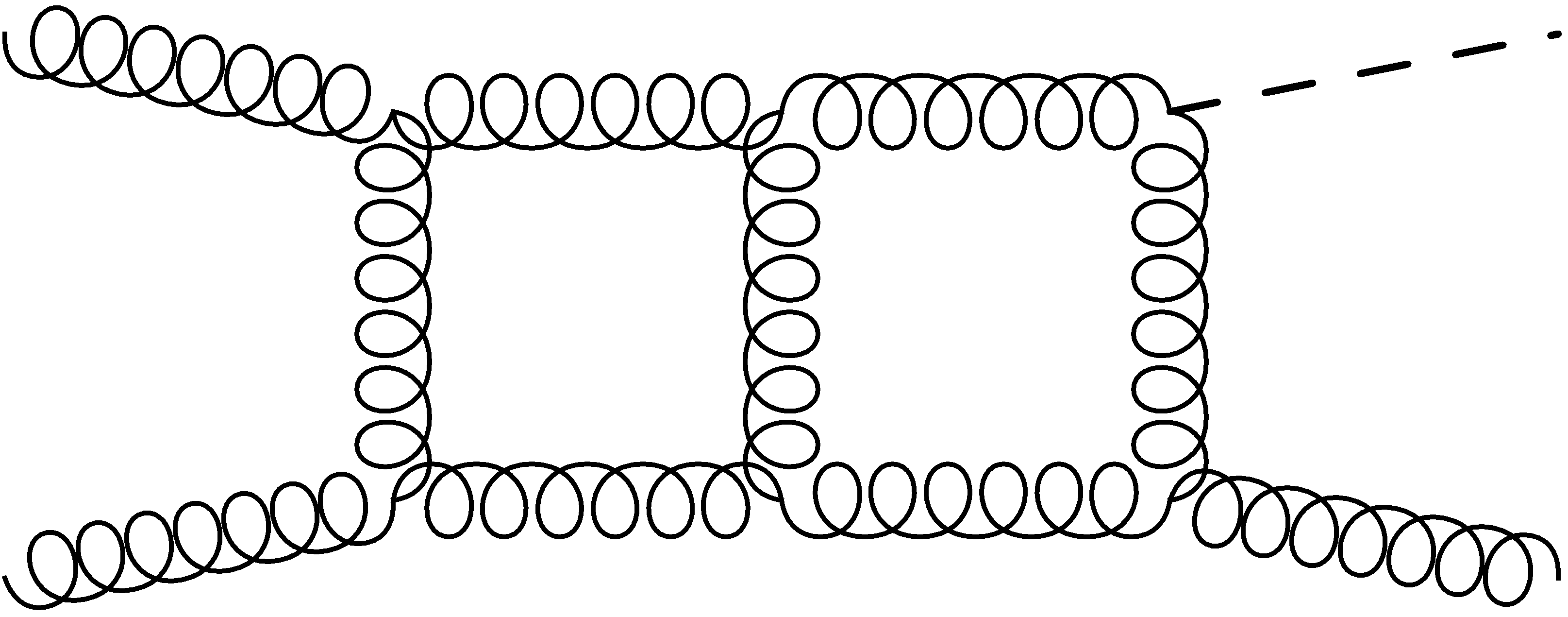} % Adjust scale as needed
\caption{Virtual-Virtual}
\label{fig:VV}
\end{subfigure}
\hfill % Adds horizontal space between the figures
% Second image with the same scale
\begin{subfigure}[b]{0.3\textwidth}
\centering % Center the image in the subfigure
\includegraphics[scale=0.04]{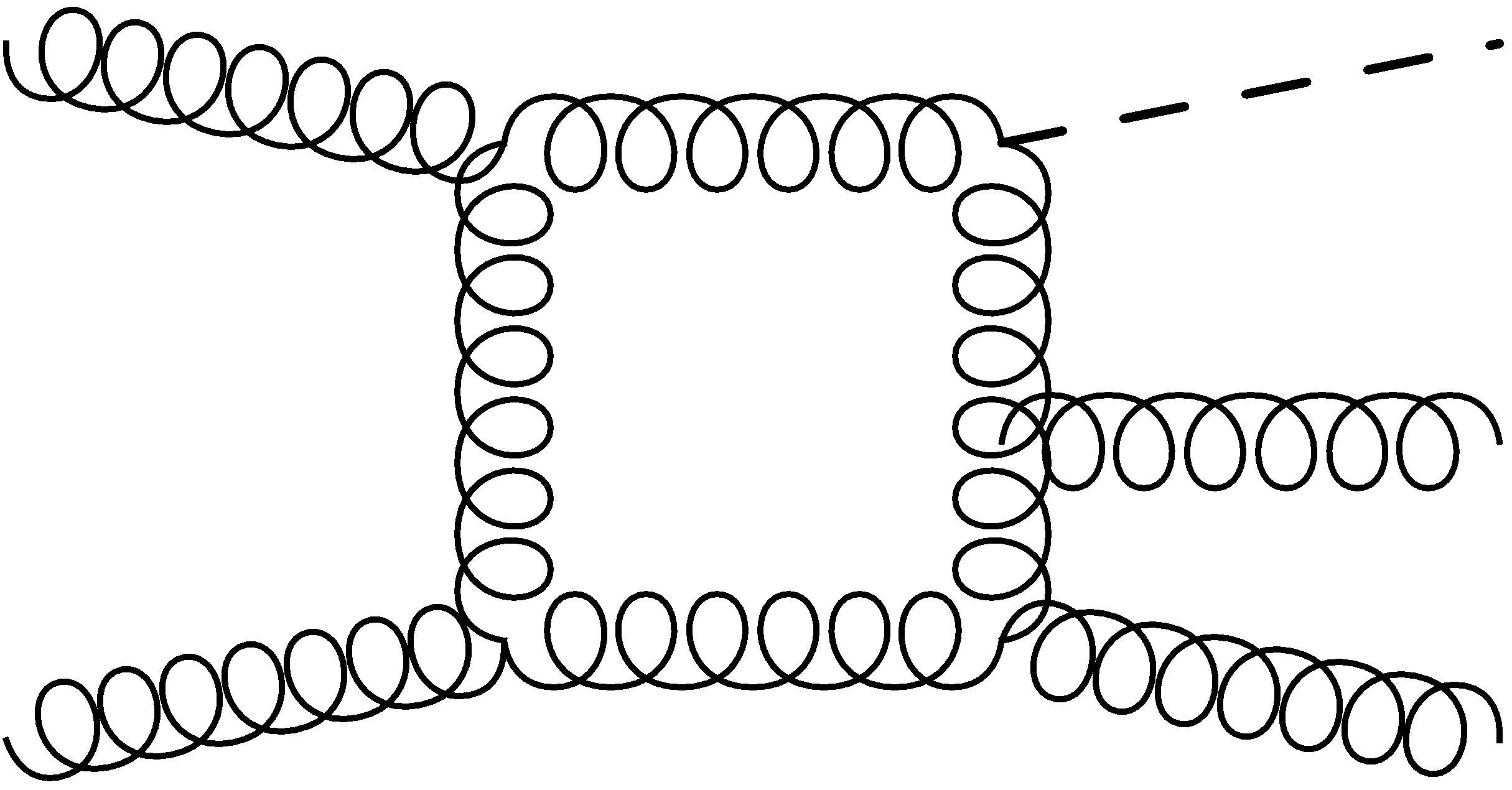} % Use the same scale
\caption{Real-Virtual}
\label{fig:RV}
\end{subfigure}
\hfill % Adds horizontal space between the figures
% Third image with the same scale
\begin{subfigure}[b]{0.3\textwidth}
\centering % Center the image in the subfigure
\includegraphics[scale=0.04]{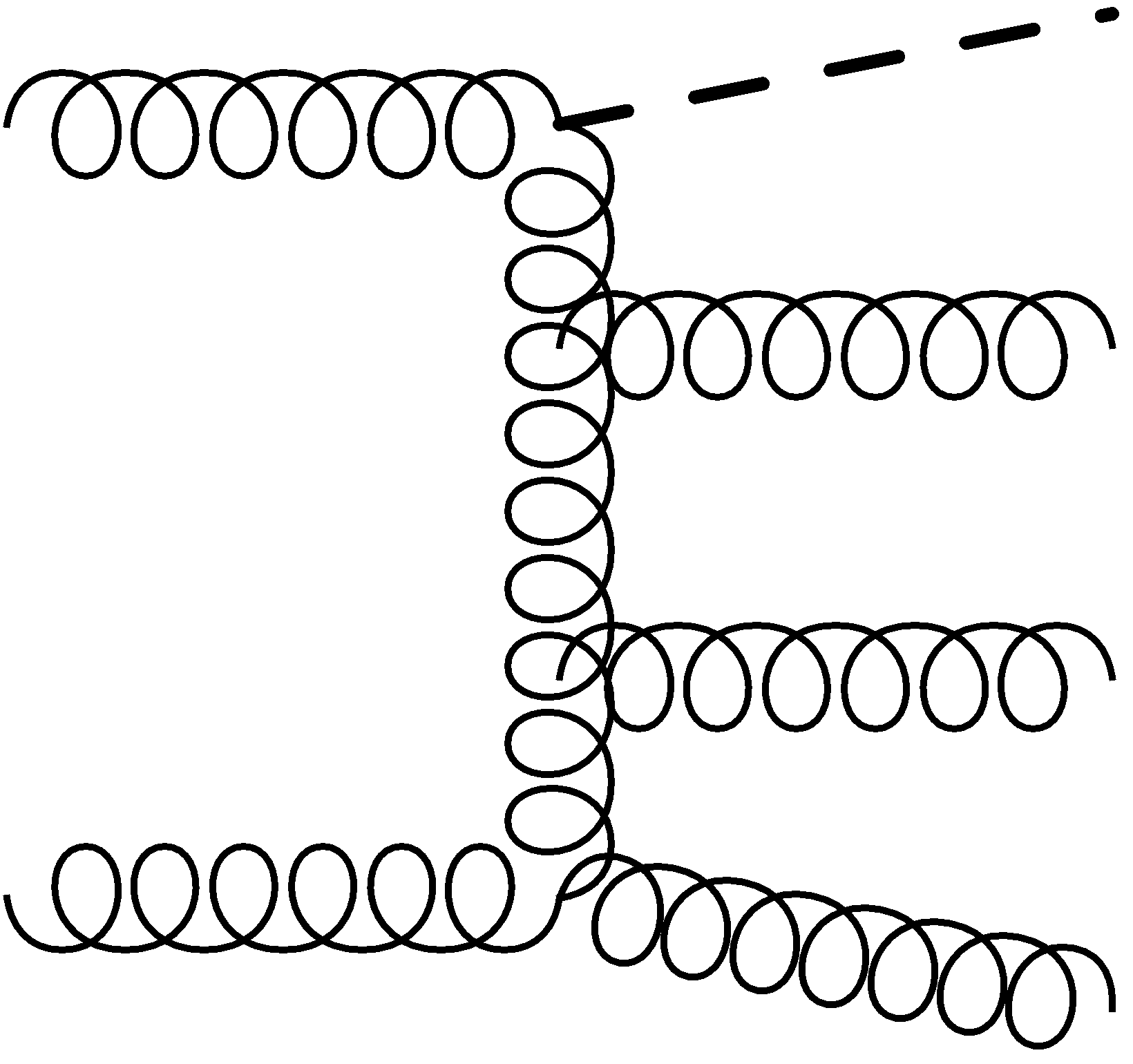} % Use the same scale
\caption{Real-Real}
\label{fig:RR}
\end{subfigure}
\caption{Representative Feynman diagrams for the production of an $A$ boson with an additional jet at NNLO.} 
\label{feynman_diagrams}
\end{center}
\end{figure}
Representative Feynman diagrams for this process at NNLO are shown in Fig.~\ref{feynman_diagrams} and Fig.~\ref{feynman_diagramsOJ}.
In each diagram the $A$ boson couples to partons through an effective Lagrangian. 
The effective Lagrangian describing the coupling of the $A$ boson to gluons (and massless quarks) arises from integrating out 
the top quark and can be written as~\cite{Chetyrkin:1998mw}\footnote{Note we factored $\frac{1}{8}$ and $\frac{1}{2}$ in the each Wilson coefficients respectively. See eq.(2.2) in~\cite{Banerjee:2017faz}.},
\begin{align} 
\mathcal{L}^{A}_{\rm eff} 
= 
- A
\Big[ 
 {C}_{G} O_{G}(x) +  {C}_{J} O_{J}(x)\Big] ,
\label{eq:effective_Lagrangian}
\end{align}
where the operators are defined as
\begin{equation}
  O_{G}(x) = G^{\mu\nu}_a \tilde{G}_{a,\mu
    \nu} \equiv  \epsilon_{\mu \nu \rho \sigma} G^{\mu\nu}_a G^{\rho
    \sigma}_a\, ,\qquad
  O_{J}(x) = \partial_{\mu} \left( \bar{\psi}
    \gamma^{\mu}\gamma_5 \psi \right)  \,.
  \label{eq:operators}
\end{equation}
The Wilson coefficients $C_G$ and $C_J$ are obtained by integrating
out the top quark loops and are defined as follows:  
\begin{align}
\label{eq:const}
& C_{G} 
= 
- \frac{\alpha_s}{2 \pi} \frac{1}{v} \left( \frac{1}{8} \right),\ 
\text{when} \ \tan{\beta} \sim 1
\\
& C_{J} 
= 
- \left[ 
\left( \frac{\alpha_s}{2 \pi} \right)  \frac{C_{F}}{4} 
            \left( \frac{3}{2} - 3\ln \frac{\mu_{R}^{2}}{m_{t}^{2}} \right) 
    + \left( \frac{\alpha_{s}}{2\pi} \right)^2 C_J^{(2)} + \cdots \right] 
C_{G}\, .
\end{align}
The Adler-Bardeen theorem~\cite{Adler:1969gk} constraints the higher order corrections to $C_G$, 
in our calculation we need $C_J$ expanded to $O(\alpha_s^2)$. The $C_{J}$ operator is one order higher in $\alpha_s$ than $C_{G}$ and as such the contributions from this operator are simpler and correspond to one-loop boxes.
The full complexities of the NNLO calculation are therefore felt in the $C_G$ piece of the calculation. It is interesting to note that the tree-level contribution from the $C_J$ pieces (i.e. the right hand diagram in Fig.~{\ref{feynman_diagramsOJ}} without the loop gluon) enter at $\mathcal{O}(\epsilon)$. Therefore as $\epsilon\rightarrow 0$ there is no contribution from $C_J$ at $\mathcal{O}(\alpha_s^4)$, but these pieces do play a role in correctly performing the renormalization at NNLO.

The topologies presented in Fig.~\ref{feynman_diagrams} are examples of double-virtual (2-loop), real-virtual (1-loop) and real-real (tree-level)
corrections which appear at $\mathcal{O}(\alpha_s^2)$ in perturbation theory. Each topology individually contains copious infra-red (IR) singularities 
which cancel only upon combinations of the sub-topologies into IR-safe observables. Over the last decade significant progress has been achieved on 
regulating IR divergences. 
In our calculation we will employ the $N$-jettiness slicing method~\cite{Gaunt:2015pea,Boughezal:2015dva}. This proceeds by defining the 1-jettiness observable as follows:
\begin{eqnarray}
\tau_1 = \sum_{j=1,m} \min_{i} \left\{ \frac{2 q_i \cdot p_{j}}{Q_i}\right\} \, ,
\end{eqnarray}
where the index $j$ runs over the momenta of all partons $p_j$ while $q_i$ represents a momenta of the 
jet-clustered phase space.
For small values of $\tau_1$ (defined by $\taucm$) a factorization theorem exists from Soft-Collinear-Effective-Theory (SCET)
~\cite{Stewart:2010tn,Stewart:2009yx}:
\begin{eqnarray}
\sigma(\tau_1 \le \taucm ) = \int_0^{\taucm}  d\tau_1 \; \left(\mathcal{S} \otimes   \mathcal{J}  \otimes  \prod_{a=1,2} \mathcal{B}_a   \otimes \mathcal{H}\right) + \mathcal{F}( \taucm ),
\label{eq:scettau}
\end{eqnarray}
The above equation is valid up to power corrections (denoted by the $ \mathcal{F}(\taucm )$ term), which vanish in the limit $\taucm  \rightarrow 0$. 
At NLO the leading power corrections are well described by the form $\taucm  \log  (\taucm /Q)$, and at NNLO the leading power corrections have the form  $\taucm  \log^3 (\taucm /Q_i)$ (where in both cases $Q_i$ is a hard 
scale associated with the process). The factorization theorem contains the universal soft ($\mathcal{S}$), jet  ($\mathcal{J}$), and beam ($\mathcal{B}$) functions, for which calculations accurate to $\mathcal{O}(\alpha_s^2)$ needed for our calculation can be found in  Refs.~\cite{Campbell:2017hsw,Boughezal:2015eha,Becher:2010pd,Becher:2006qw,Gaunt:2014xga,Gaunt:2014cfa}. Specifically we use the implementation in MCFM described in Refs.~\cite{Boughezal:2016wmq,Campbell:2019dru}. 
The process-specific hard function $\mathcal{H}$ must be calculated on a case-by-case basis and is obtained from the double-virtual type topologies in Fig.~\ref{feynman_diagrams}.In our calculation we can therefore recycle much of the results presented for $H+j$ at NNLO~\cite{Campbell:2019gmd}, which given the similarity to our final state of interest, differs only by the hard function. We describe the calculation of the hard function in detail at $\mathcal{O}(\alpha_s^2)$ in section~\ref{sec:hardfunction}.
Finally, we note that there are various options regarding the choice of the hard scale $Q_i$, appearing in the definitions above. 
The most common choice of $Q_i$ is $Q_i = 2E_i$ which is the so-called geometric measure~\cite{Jouttenus:2011wh,Jouttenus:2013hs}. 
However in Ref.~\cite{Campbell:2019gmd}, for higher order corrections to Higgs plus jet it was shown that a boosted frame, in which the final-state system is at rest, resulted in a smaller impact of power corrections. 
We also have confirmed this boosted choice shows better asymptotic behavior compared to
the geometric measure, and therefore will present results using the boosted definition in this paper.

The SCET-factorization theorem describes the cross section in region below the cut, leaving the above-cut region $\tau_1 > \taucm$ to be determined. 
The primary advantage of the slicing method resides in the fact that in the above-cut region there is at least one resolved emission, and therefore the phase space contains only NLO IR-singularities. Therefore in order to describe this region, a NLO calculation of $A+2j$ is all that is required. We discuss this calculation in detail in section~\ref{sec:abovecut}.

%%%%%%%%%%%%%%%%%%%%%%%%%%twoloop calculation%%%%%%%%%%%%%%%%%%%%%%%%%%%
\section{Hard function for $A\rightarrow ggg$ and $A\rightarrow q\overline{q}g$ at NNLO}
\label{sec:hardfunction}

\subsection{Overview}
In this section we describe the calculation of the hard function for the process $A\rightarrow ggg$ and $A\rightarrow q\overline{q}g$ at NNLO accuracy ($\mathcal{O}(\alpha_s^5)$) in the $m_t \rightarrow \infty$ limit.
We begin by considering the decay of an $A$ boson to partons. At LO these reactions can be denoted as follows: 
\begin{align*}
A \to g(p_1)\,g(p_2)\,g(p_3)  \quad {\rm{and}} \quad\ A \to q(p_1)\,\bar{q}(p_2)\,g(p_3) \, .
\end{align*}
We define Mandelstam invariants for this process as
\begin{align*}
s = (p_1+p_2)^2 > 0 \, , \hspace{1.5cm} t = (p_1+p_3)^2 > 0 \, , \hspace{1.5cm} u = (p_2+p_3)^2 > 0 \, ,
\end{align*}
and from momentum conservation these satisfy $s+t+u=m_A^2$, where $m_A$ denotes the mass of the pseudoscalar boson. We also introduce the dimensionless quantities
\begin{align}
x = \frac{s}{m_A^2} \, , \hspace{1cm} y = \frac{t}{m_A^2} \, , \hspace{1cm} z = \frac{u}{m_A^2} \, ,
\end{align}
which satisfy $0<x<1$, $0<y<1$, $0<z<1$, and $x+y+z=1$ (for decay kinematics).
In order to obtain predictions for the collider production of an $A$ boson with an additional jet, the decay amplitudes must be crossed to generate the relevant partons in the initial state, this alters the allowed regions in $(x,y,z)$ space, which we will discuss shortly. We note that the calculation for the decay kinematics $A\rightarrow$ 3 partons has been considered in the literature~\cite{Banerjee:2017faz}. Although to allow for an efficient mechanism of generating the crossed process we have re-calculated the 2-loop amplitudes independently. However, where possible we try to maintain the notation from~\cite{Banerjee:2017faz} to facilitate an easy comparison, whenever we deviate from the notation in the literature we will mention it\footnote{For instance, we note that $t \leftrightarrow u$ and $y \leftrightarrow z$
in~\cite{Banerjee:2017faz}.}.

%%%%%%%%%%%%%%%%%%%%%%%%%%twoloop calculation%%%%%%%%%%%%%%%%%%%%%%%%%%%
%\subsection{Lagrangian, kinematics, and notation}
%We consider the processes 

\subsection{Calculation}

We now discuss the calculation of the amplitudes up to two-loop order. 
The effective Lagrangian contains both $\epsilon_{\mu \nu \rho \sigma}$ and $\gamma_5$ which are inherently four-dimensional objects. Care must therefore be taken when calculating using the dimensional regulator $d=4-2\epsilon$. We proceed as follows, $\gamma_5$ is defined as
%the Hooft-Veltman prescription~\cite{tHooft:1972tcz} which is one of the most popular ways and is compatible with the CDR calculation. The prescription defines each as follows
%
\begin{align}
  \gamma_5 = \frac{i}{4!} \varepsilon_{\mu \nu \rho \sigma}
  \gamma^{\mu}  \gamma^{\nu} \gamma^{\rho} \gamma^{\sigma} \,,
\end{align}
\begin{align}
  \label{eqn:LeviContract}
  \varepsilon_{\mu_1\nu_1\rho_1\sigma_1}\,\varepsilon^{\mu_2\nu_2\rho_2\sigma_2}=
  \large{\left |
  \begin{array}{cccc}
    \delta_{\mu_1}^{\mu_2} &\delta_{\mu_1}^{\nu_2}&\delta_{\mu_1}^{\rho_2} & \delta_{\mu_1}^{\sigma_2}\\
    \delta_{\nu_1}^{\mu_2}&\delta_{\nu_1}^{\nu_2}&\delta_{\nu_1}^{\rho_2}&\delta_{\nu_1}^{\sigma_2}\\
    \delta_{\rho_1}^{\mu_2}&\delta_{\rho_1}^{\nu_2}&\delta_{\rho_1}^{\rho_2}&\delta_{\rho_1}^{\sigma_2}\\
    \delta_{\sigma_1}^{\mu_2}&\delta_{\sigma_1}^{\nu_2}&\delta_{\sigma_1}^{\rho_2}&\delta_{\sigma_1}^{\sigma_2}
  \end{array}
  \right |}
\end{align}
where all the Lorentz indices are $d$-dimensional~\cite{Larin:1993tq}. This is commonly referred to as the Larin prescription.
While being straightforward to implement in $d$ dimensions, the above definition fails to satisfy the Ward identities, and therefore an additional, finite, renormalization of $\gamma_5$ is required in order to restore the Ward Identity. 
%which causes issues the UV-renormalization of the amplitudes due to the additional finite terms to restore the axial renormalization current~\cite{Larin:1993tq}.

Aside from the issues with $\gamma_5$ there are additional complications arising from operator mixing. The essential details of the UV renormalization procedure are well summarized in Refs.~\cite{Ahmed:2015qpa,Banerjee:2017faz}, here for brevity we provide a brief overview and we refer the reader to the literature for further details. We renormalize the bare strong coupling constant by performing the replacement
\begin{equation}
\hat{\alpha}_s \to \as\, S_{\epsilon} \, Z_{\alpha} \label{asren}  \, ,  
\end{equation}
with $S_{\epsilon} = \frac{\exp{(\epsilon \gamma_E)}}{(4\pi)^{\epsilon}}$, $\as \equiv \as(\mu_R)$ at the renormalization scale $\mu_R$ which keeps  the coupling constant dimensionless in $d=4-2\epsilon$. The renormalization factors are given by
\begin{align}
Z_{\alpha} &= 1 + \left (\frac{\as}{2\pi}\right ) r_1 + \left (\frac{\as}{2\pi}\right )^2 r_2 + \mathcal{O}(\alpha_s^3) \, ,
\label{eq:Zalphas}
\end{align}
where explicit formula for the terms $r_1$, $r_2$ are defined in Appendix \ref{renoirformulae}. After coupling renormalization the sub-amplitudes for the two operators $O_{\Lambda=G}$ and $O_{\Lambda=J}$ for each process $f=ggg,\ q\bar{q}g$ are written as,
\begin{align}
\mathcal{M}_f^{\Sigma,(0)} &= \hat{\mathcal{M}}_f^{\Sigma,(0)} \, , \\
\mathcal{M}_f^{\Sigma,(1)} &= S_{\epsilon} \hat{\mathcal{M}}_f^{\Sigma,(1)} 
+\frac{r_1}{2} \hat{\mathcal{M}}_f^{\Sigma,(0)} \, , \\
\mathcal{M}_f^{\Sigma,(2)} &= S_{\epsilon}^2 \hat{\mathcal{M}}_f^{\Sigma,(2)}
+\frac{3 r_1}{2} S_{\epsilon} \hat{\mathcal{M}}_f^{\Sigma,(1)} \notag \\ &\quad 
+\left( \frac{r_2}{2} -\frac{r_1^2}{8} \right) \hat{\mathcal{M}}_f^{\Sigma,(0)} \, ,
\end{align}
where $\Sigma = G, J$.
%Afterwards, the UV-finite amplitudes with the additional operator renormalization with mixing terms can be written as 
Next, applying the renormalization of $\gamma_5$ and operator mixing described previously we define
\begin{eqnarray}
\mathcal{M}_f^{\Lambda} = \sum_{\Sigma=G,J} Z_{\Lambda \Sigma}
\left( 
\mathcal{M}_f^{\Sigma,(0)} 
+ \left( \frac{\alpha_s}{2\pi} \right) \mathcal{M}_f^{\Sigma,(1)} 
+ \left( \frac{\alpha_s}{2\pi} \right)^2 \mathcal{M}_f^{\Sigma,(2)} 
+ \mathcal{O}(\alpha_s^3)
\right),
\label{eq:UV_fin_amp}
\end{eqnarray}
all the necessary renormalization coefficients can be found in~\cite{Larin:1993tq,Zoller:2013ixa} and we just simply present them with our notation here,
\begin{align}
Z_{G G} &= 1 + \left (\frac{\as}{2\pi}\right ) z_{GG,1} 
+ \left (\frac{\as}{2\pi}\right )^2 z_{GG,2} + \mathcal{O}(\alpha_s^3) 
\label{eq:Z_GG} \\
Z_{G J} &= \left (\frac{\as}{2\pi}\right ) z_{GJ,1} 
+ \left (\frac{\as}{2\pi}\right )^2 z_{GJ,2} + \mathcal{O}(\alpha_s^3) 
\label{eq:Z_GJ} \\
Z_{J G} &= 0  
\label{eq:Z_JG} \\
Z_{J J} &= 1 + \left (\frac{\as}{2\pi}\right ) z_{JJ,1} 
+ \left (\frac{\as}{2\pi}\right )^2 z_{JJ,2} + \mathcal{O}(\alpha_s^3),
\label{eq:Z_JJ} 
\end{align}
where $z_{\Lambda, \Sigma}$ can be found in Appendix \ref{renoirformulae}. 
After applying the prescription defined above we finally obtain UV-finite amplitudes.
These amplitudes can be written as 
\begin{align}
\mathcal{A}_f = \left( 4 \pi \alpha_s \right)^{1/2} \sum_{\Lambda=G,J} 
C_{\Lambda}(\alpha_s) \mathcal{M}_f^{\Lambda} .
\label{eq:amp_expand}
\end{align}
Such that, with ~\eqref{eq:UV_fin_amp} and ~\eqref{eq:amp_expand}, we obtain all the necessary ingredients to define the squared amplitudes.
For $f=ggg$,
\begin{align}
\mathcal{A}_{ggg} \mathcal{A}_{ggg}^* 
&= (4 \pi \alpha_s) \left( \frac{\alpha_s}{2 \pi} \right)^2 \left( C_G^{(1)} \right)^2
\Bigg[  
\mathcal{M}_{ggg}^{G,(0)} \mathcal{M}_{ggg}^{G,(0)*} 
+ \left( \frac{\alpha_s}{2 \pi} \right) 2 \,\text{Re}\left(  \mathcal{M}_{ggg}^{G,(1)} \mathcal{M}_{ggg}^{G,(0)*}  \right) \notag \\
&+ \left( \frac{\alpha_s}{2 \pi} \right)^2 
\bigg[
2 \,\text{Re}\left(  \mathcal{M}_{ggg}^{G,(2)} \mathcal{M}_{ggg}^{G,(0)*} \right) 
+ \mathcal{M}_{ggg}^{G,(1)} \mathcal{M}_{ggg}^{G,(1)*} 
+ 2 C_J^{(1)} \,\text{Re} \left( \mathcal{M}_{ggg}^{J,(1)} \mathcal{M}_{ggg}^{G,(0)*} \right)  
\bigg]
\Bigg] 
\label{eq:ggg_amp_sq}
\end{align}
and for $f=q\bar{q} g$,
\begin{align}
\mathcal{A}_{q\bar{q}g} \mathcal{A}_{q\bar{q}g}^* 
&= (4 \pi \alpha_s) \left( \frac{\alpha_s}{2 \pi} \right)^2 \left( C_G^{(1)} \right)^2
\Bigg[
\mathcal{M}_{q\bar{q}g}^{G,(0)} \mathcal{M}_{q\bar{q}g}^{G,(0)*} \notag \\
&+ \left( \frac{\alpha_s}{2 \pi} \right)
\bigg[
2 \,\text{Re}\left(  \mathcal{M}_{q\bar{q}g}^{G,(1)} \mathcal{M}_{q\bar{q}g}^{G,(0)*} \right)
+ 2 C_J^{(1)} \mathcal{M}_{q\bar{q}g}^{G,(0)} \mathcal{M}_{q\bar{q}g}^{J,(0)*}
\bigg] \notag \\
&+ \left( \frac{\alpha_s}{2 \pi} \right)^2
\bigg[
2 \,\text{Re}\left( \mathcal{M}_{q\bar{q}g}^{G,(2)} \mathcal{M}_{q\bar{q}g}^{G,(0)*} \right)
+ \mathcal{M}_{q\bar{q}g}^{G,(1)} \mathcal{M}_{q\bar{q}g}^{G,(1)*} \notag \\
&+ 2 C_J^{(1)} \left( \mathcal{M}_{q\bar{q}g}^{G,(1)} \mathcal{M}_{q\bar{q}g}^{J,(0)*}
+ \mathcal{M}_{q\bar{q}g}^{G,(0)} \mathcal{M}_{q\bar{q}g}^{J,(1)*} \right) \notag \\
&+ \left( C_J^{(1)} \right)^2 \mathcal{M}_{q\bar{q}g}^{J,(0)} \mathcal{M}_{q\bar{q}g}^{J,(0)*}
+ 2 C_J^{(2)} \mathcal{M}_{q\bar{q}g}^{G,(0)} \mathcal{M}_{q\bar{q}g}^{J,(0)*}  
\bigg]
\Bigg] \  ,
\label{eq:qqg_amp_sq}
\end{align}
where
\begin{align}
C_{G}^{(1)} = - \frac{1}{v} \left( \frac{1}{8} \right) \,,
\quad \quad
C_{J}^{(1)} = - \frac{C_{F}}{4} \left( \frac{3}{2} - 3\ln \frac{\mu_{R}^{2}}{m_{t}^{2}} \right)   .
\end{align}
The sum over the polarization states of the external gluons is defined by the completeness relation 
\begin{align}
\sum_{\text{pol}} \epsilon^{\mu}(p_i) \epsilon^{\nu *}(p_i) = -g^{\mu \nu} + \frac{p_i^{\mu} q^{\nu} + q^{\mu} p_i^{\nu}}{q \cdot p_i}
\end{align}
where $q$ is an auxiliary vector. In our calculation we choose $q=p_2, p_3, p_1$ for $p_i=p_1,p_2,p_3$ respectively.\\
\\
The workflow of our calculation proceeds in the following manner. 
Initially we generated the tree-level, one-loop, and two-loop Feynman diagrams using FeynArts~\cite{Hahn:2000kx}. We confirmed the results using an independent implementation using QGRAF~\cite{Nogueira:1991ex}. For the FeynArts model file generation we have used FeynRules~\cite{Christensen:2008py,Alloul:2013bka}. 
The Feynman rules have been implemented using FeynCalc~\cite{Mertig:1990an,Shtabovenko:2020gxv}.
The resulting one-loop and two-loop amplitudes can be written in terms of scalar loop integrals. These integrals are then further reduced to a smaller set of Master Integrals (MIs). The reduction to MI's was performed using LiteRed~\cite{Lee:2013mka} and KIRA~\cite{Maierhofer:2017gsa}.
At the one-loop level all the integrals are reduced to the bubble and box integral which can be found in Appendix A of Ref.~\cite{Garland:2001tf}. At the two-loop level all integrals can be expressed in terms of MIs which occur in the similar $H\rightarrow$ 3 partons process,  which are presented in Ref.~\cite{Gehrmann:2000zt,Gehrmann:2001ck}. All of the MIs are expressed in terms of 
HPLs \cite{Remiddi:1999ew} and two-dimensional HPLs (2dHPLs) \cite{Gehrmann:2000zt,Gehrmann:2001ck}.
We also use GiNaC~\cite{Bauer:2000cp} and HandyG~\cite{Naterop:2019xaf} to numerically evaluate the HPLs and 2dHPLs for checking purposes.

As a cross check of our calculation we calculated both the desired $A\rightarrow 3$ parton process and $H\rightarrow 3$ partons, the later of which has been studied extensively in the literature. We also checked the various crossings of our amplitudes to the physical kinematic regions in the same way. 
While the (re-)calculation of $H+j$ is not the primary focus of this paper or discussion herein  we provide a detailed summary of our Higgs calculation (and comparison to the literature) in Appendix~\ref{sec:Higgs_recalculation}.

%%%%%%%%%%%%%%%%%%%%%%%%%%%%%%%%%%%%%%%%%%%%%%%%%%%%%%%%%%%%%%%%%%%%%%%%%%%%%%%%%%%%%%%%%%%%%%%%%%%%%%%%%%%%
\subsection{IR subtraction and conversion to $\overline{\text{MS}}$ scheme}

In order to define the hard function utilized in the SCET factorization theorem, the amplitudes must be converted to a suitable form. In order to obtain the hard function we remove the explicit soft and collinear divergences from the UV-renormalized coefficients. The IR structure of one-loop and two-loop QCD amplitudes is universally known \cite{Catani:1998bh} and can be written using Catani's subtraction operators 
$\bI^{(\ell)}(\epsilon)$. 
The finite coefficients $\mathcal{M}_f^{\Lambda,\text{fin}}$ are defined as
\begin{align}
\mathcal{M}_f^{\Lambda,(0),\text{fin}} &= \mathcal{M}_f^{\Lambda,(0)} \, , \\
\mathcal{M}_f^{\Lambda,(1),\text{fin}} &= \mathcal{M}_f^{\Lambda,(1)} - \bI_f^{(1)}(\epsilon) \mathcal{M}_f^{\Lambda,(0)} \, , \\
\mathcal{M}_f^{\Lambda,(2),\text{fin}} &= \mathcal{M}_f^{\Lambda,(2)} 
- \bI_f^{(1)}(\epsilon) \mathcal{M}_f^{\Lambda,(1)} 
- \bI_f^{(2)}(\epsilon) \mathcal{M}_f^{\Lambda,(0)} \, .
\end{align}
We confirmed the IR-singularities in our amplitude cancel after this operation as desired. 
The explicit expressions of the subtraction operators for $A\rightarrow ggg, q\bar{g}g$ can be found in Appendix \ref{renoirformulae}. 
In Appendix \ref{app:NLO_results}, we provide full results for the tree and one-loop expressions 
for $\mathcal{M}_f^{\Lambda,\text{fin}}$ for a quick comparison for readers. 
The full results for the two-loop matrix elements can be obtained from the attached ancillary files with the {\tt{arXiv}} submission.
The description of the files can be found in Appendix \ref{app:ancillary_files}. \\

Finally, following the discussion in Ref.~\cite{Becher:2013vva}, we obtain the $\overline{\text{MS}}$-renormalized coefficients $\mathcal{M}_f^{\Lambda,\text{ren}}$ in the following way:
\begin{align}
\mathcal{M}_f^{\Lambda,(0),\text{ren}} &= \mathcal{M}_f^{\Lambda,(0),\text{fin}} \, , \\
\mathcal{M}_f^{\Lambda,(1),\text{ren}} &= \mathcal{M}_f^{\Lambda,(1),\text{fin}} 
+ \left( \bm{I}_f^{(1)}(\epsilon) + \bm{Z}_f^{(1)}(\epsilon) \right) \mathcal{M}_f^{\Lambda,(0),\text{fin}} \label{mscoeff1} \, , \\
\mathcal{M}_f^{\Lambda,(2),\text{ren}} &= \mathcal{M}_f^{\Lambda,(2),\text{fin}} 
+ \left( \bm{I}_f^{(1)}(\epsilon) + \bm{Z}_f^{(1)}(\epsilon) \right) \mathcal{M}_f^{\Lambda,(1),\text{fin}} 
\notag \\
&+ \Bigg[ 
\bI_f^{(2)}(\epsilon) 
+ \left(\bI_f^{(1)}(\epsilon) +\boldsymbol{Z}_f^{(1)}(\epsilon)\right) \bI_f^{(1)}(\epsilon) 
+ \boldsymbol{Z}_f^{(2)}(\epsilon)\Bigg] 
\mathcal{M}_f^{\Lambda,(0),\text{fin}} \, ,
\label{mscoeff2}
\end{align}
where $\boldsymbol{Z}_f^{(1)}$ and $\boldsymbol{Z}_f^{(2)}$ for $f=ggg, q\bar{q}g$ are presented in Appendix \ref{renoirformulae}.

As mentioned previously, in order to evaluate the amplitude in a kinematic region relevant for scattering at the LHC the amplitudes must be crossed to ensure the correct partons are in the initial state. 
The detailed procedure on how to achieve this can be found in Refs.~\cite{Gehrmann:2011ab,Gehrmann:2002zr}. In this article, on the other hand, we followed coproduct method~\cite{Duhr:2014woa,Duhr:2012fh} as outlined in Ref.~\cite{Mondini:2019vub}.
We have confirmed the consistency of our crossed MI's using a numerical evaluation of the MI's in the scattering region using the package AMFlow~\cite{Liu:2022chg}.
\\

In the following we present numerical results of hard functions at each channel using the parameters defined in Eq.\eqref{eq:Hard_inputs} with $m_A = 0.1\,\text{TeV}$:
\begin{itemize}
\item $gg$-channel:
\begin{align}
\label{eq:hard_A_num_res_gg}
\text{Re}\left(  \mathcal{M}_{gg\to A g}^{G,\text{ren}} \, \mathcal{M}_{gg\to A g}^{G*,\text{ren}} \right) 
&=
4.85921 + (38.5429) \, \alpha _s \nonumber\\ &+ (187.3091)  \,\alpha _s^2 \,,
\notag \\
\text{Re}\left(  
\mathcal{M}_{gg\to A g}^{J,\text{ren}} \, \mathcal{M}_{gg\to A g}^{G*,\text{ren}} 
+
\mathcal{M}_{gg\to A g}^{G,\text{ren}} \, \mathcal{M}_{gg\to A g}^{J*,\text{ren}} 
\right) 
&=
0 + (0) \, \alpha _s + (1.23085)  \,\alpha _s^2 \,,
\notag \\
\end{align}

\item $q\bar{q}$-channel:
\begin{align}
\label{eq:hard_A_num_res_qqb}
\text{Re}\left(  
\mathcal{M}_{q \bar{q}\to A g}^{G,\text{ren}} \, 
\mathcal{M}_{q \bar{q}\to A g}^{G*,\text{ren}}
\right) 
&=
0.50810 + (1.38529) \, \alpha _s  \nonumber\\ &+ (3.67559)  \,\alpha _s^2 \,,
\notag \\
\text{Re}\left(  
\mathcal{M}_{q \bar{q}\to A g}^{J,\text{ren}} \, 
\mathcal{M}_{q \bar{q}\to A g}^{G*,\text{ren}}
+
\mathcal{M}_{q \bar{q}\to A g}^{G,\text{ren}} \, 
\mathcal{M}_{q \bar{q}\to A g}^{J*,\text{ren}}
\right) 
&=
0 + (0) \, \alpha _s + (0.12870)  \,\alpha _s^2 \,,
\notag \\
\text{Re}\left(  
\mathcal{M}_{q \bar{q}\to A g}^{J,\text{ren}} \, 
\mathcal{M}_{q \bar{q}\to A g}^{J*,\text{ren}}
\right) 
&=
(0) + (0) \, \alpha _s + (0)  \,\alpha _s^2 \,,
\notag \\
\end{align}

\item $qg$-channel:
\begin{align}
\label{eq:hard_A_num_res_qg}
\text{Re}\left(  \mathcal{M}_{q g\to A g}^{G,\text{ren}} 
\, \mathcal{M}_{q g\to A g}^{G*,\text{ren}} \right) 
&=
-1.96610 + (-16.4889) \, \alpha _s  \nonumber\\ &+ (-77.2028)  \,\alpha _s^2 \,,
\notag \\
\text{Re}\left(  
\mathcal{M}_{q g\to A g}^{J,\text{ren}} 
\, \mathcal{M}_{q g\to A g}^{G*,\text{ren}} 
+
\mathcal{M}_{q g\to A g}^{G,\text{ren}} 
\, \mathcal{M}_{q g\to A g}^{J*,\text{ren}} 
\right) 
&=
0 + (0) \, \alpha _s + (-0.49802)  \,\alpha _s^2 \,,
\notag \\
\text{Re}\left(  \mathcal{M}_{q g\to A g}^{J,\text{ren}} 
\, \mathcal{M}_{q g\to A g}^{J*,\text{ren}} \right) 
&=
(0) + (0) \, \alpha _s + (0)  \,\alpha _s^2 \,.
\end{align}

\end{itemize}
We note that the above results have not been normalized by the leading order results. 

\subsection{Factorization properties of the two-loop amplitude}

As this section has detailed, the calculation of the second order hard function is quite intricate and involves several distinct stages. Through diagram generation, reduction to MI's, the definition of a UV finite amplitude, extraction of IR poles, evaluation and analytically continuing the decay amplitudes to the scattering region. Therefore it is natural to search for methods of validating the correctness of our approach. In this section we outline the checks which we have performed on our result, which correspond to testing the IR singular limits of our calculation against the know factorization properties of QCD~\cite{Badger:2004uk,Gehrmann:2011ab}. 

A particularly good test is the collinear limit~\cite{Badger:2004uk}, we consider the two cases for $f=ggg,q\bar{q}g$ separately. For the $f=ggg$ case, we consider a phase space point in which one of the gluons ($g=g(p_3)$) becomes collinear to an another gluon ($g=g(p_1)$), as a result of which the invariant $t$ vanishes which means $y \to 0$ while $x,z\neq 0$.
For the $f=q\bar{q}g$ case, we consider phase space point where the $q\overline{q}$ pair becomes collinear, and as such the invariant $s$ vanishes which means $x \to 0$ while $y,z\neq 0$.
The collinear limit at two loops reads:
\begin{align}
\hat{\mathcal{M}}^{G,(2)}_f \hat{\mathcal{M}}^{G,(0)*}_f
&\to 
C^{(2)}_f = 
P_f^{(0)} \cdot \hat{\mathcal{M}}^{G,(2)}_{A \to gg} \hat{\mathcal{M}}^{G,(0)*}_{A \to gg} \notag \\ 
&\qquad + P_f^{(1)} \cdot \hat{\mathcal{M}}^{G,(1)}_{A \to gg} \hat{\mathcal{M}}^{G,(0)*}_{A \to gg} \notag \\
&\qquad + P_f^{(2)} \cdot \hat{\mathcal{M}}^{G,(0)}_{A \to gg} \hat{\mathcal{M}}^{G,(0)*}_{A \to gg}  \, . 
\label{collinearlimiteq}
\end{align}
The splitting functions $P_f^{(\ell)}(y,z)$ and required amplitudes
$\mathcal{M}^{G,(\ell)}_{A \to gg} \mathcal{M}^{G,(0)*}_{A \to gg}$ for $\ell = 0,1,2$ are given in Appendix \ref{matrixelements}. 
We compared our result for 
$\hat{\mathcal{M}}^{G,(2)}_f \hat{\mathcal{M}}^{G,(0)*}_f$ 
as a series in $\epsilon$ with $C^{(2)}_f$. 
We multiply both expressions by a factor of $x$ or $y$ to remove the leading divergence. The numerical results are displayed in table \ref{tab1:collinear}. We observe excellent agreement between our result and the known collinear limit.
Additionally we also investigated the soft limit $p_3\rightarrow 0$ for $f=ggg$ by following the procedure outlined in section 8. of Ref.~\cite{Badger:2004uk} finding excellent agreement, we note this check is much more intricate for $f=q\bar{q}g$ case due to the effective operator structure, so we did not pursue it here.

\begin{table}
\centering
\begin{tabular}{|c|c|c|c|c|}
\hline
\rule{0pt}{2ex}{Coefficient} & 
$y\,C^{(2)}_{ggg}$ & 
$y\,\hat{\mathcal{M}}^{G,(2)}_{ggg} \hat{\mathcal{M}}^{G,(0)*}_{ggg}$ &
$x\,C^{(2)}_{q\bar{q}g}$&
$x\,\hat{\mathcal{M}}^{G,(2)}_{q\bar{q}g} \hat{\mathcal{M}}^{G,(0)*}_{q\bar{q}g}$
\\ \hline
$\epsilon^{-4}$ & $1.20981960\cdot 10^6$ & $1.20981960\cdot 10^6$ & $4.05026555\cdot 10^2$ & $4.05026555\cdot 10^2$
\\ \hline
$\epsilon^{-3}$ & $1.58228295\cdot 10^7$ & $1.58228295\cdot 10^7$ & $-2.59019027\cdot 10^3$ & $-2.59019027\cdot 10^3$
\\ \hline
$\epsilon^{-2}$ & $2.36283980\cdot 10^8$ & $2.36283980\cdot 10^8$ & $-1.20976857\cdot 10^4$ & $-1.20976857\cdot 10^4$
\\ \hline
$\epsilon^{-1}$ & $2.58965014\cdot 10^9$ & $2.58966527\cdot 10^9$ & $5.16726263\cdot 10^4$ & $5.16726262\cdot 10^4$
\\ \hline
$\epsilon^{0}$ & $2.19247701\cdot 10^{10}$ & $2.19253448\cdot 10^{10}$ & $2.38532152\cdot 10^5$ & $2.38475465\cdot 10^5$
\\ \hline
\end{tabular}
\caption{Numerical comparison between our two-loop results and the known collinear limit at phase space points 
$(y,z) = (10^{-10}, 0.23)$ and $(x,z) = (10^{-10}, 0.23)$ for $f=ggg,q\bar{q}g$ respectively with $\mu_R = m_A = 125\, \text{GeV}$. 
For readability here we have only presented the real part, although we note that we have found excellent agreement for the imaginary part also.}
\label{tab1:collinear}
\end{table}

%%%%%%%%%%%%%%%%%%%%%%%%%%%%%%%%%%%%%%%%%%%%%%
\subsection{Comparison with existing results}

As a further check of our results, we can compare our calculation for the squared amplitudes for $\mathcal{M}_f^{\Lambda,(\ell),\text{fin}}$ up to $\ell=2$ with the existing results in the literature \cite{Banerjee:2017faz}, for decay kinematics. 
We have performed this check and we have found that at tree- and one-loop level our two calculations are perfectly in agreement. 
The two-loop level comparisons are non-trivial. We begin by taking account of the following differences in notation;
\begin{align}
S_q^{G,(2)} \quad \text{Eq.(2.32) in~\cite{Banerjee:2017faz}}
&\leftrightarrow 
\left(\frac{1}{8}\right)^{-2} \left(\frac{1}{2}\right)^{-2}
\Bigg[
2 \,\text{Re}\left( \mathcal{M}_{q\bar{q}g}^{G,(2)} \mathcal{M}_{q\bar{q}g}^{G,(0)*} \right)
+ \mathcal{M}_{q\bar{q}g}^{G,(1)} \mathcal{M}_{q\bar{q}g}^{G,(1)*}
\Bigg]
\label{eq:qqg_comparison}
\\
S_g^{G,(2)} \quad \text{Eq.(2.30) in~\cite{Banerjee:2017faz}}
&\leftrightarrow
\left(\frac{1}{8}\right)^{-2} \left(\frac{1}{2}\right)^{-2}
\Bigg[
2 \,\text{Re}\left(  \mathcal{M}_{ggg}^{G,(2)} \mathcal{M}_{ggg}^{G,(0)*} \right) 
+ \mathcal{M}_{ggg}^{G,(1)} \mathcal{M}_{ggg}^{G,(1)*} 
\Bigg]
\label{eq:ggg_comparison}
\end{align}
where the factors $(1/8)^{-2}$ and $(1/2)^{-2}$ are due to the different definitions of the Lagrangian~\eqref{eq:effective_Lagrangian} and expansion in units of $\alpha_s / 2\pi$ versus of $\alpha_s / 4\pi$ as in~\cite{Banerjee:2017faz}.
Our results should be expanded with a factor $(-1)^{-2 \epsilon}$ before comparison due to our convention
$\mu_R^2 = m_A^2$ as opposed to the choice $\mu^2 = -m_A^2$ in the literature result.
Also, note that we normalized the amplitudes by Born color factor $C_A C_F$, $2 C_A^2 C_F$ for $f=q\bar{q}g, ggg$
respectively as in the literature. After making these adjustments we have found perfect agreement with the literature results for the case $f=q\bar{q}g$, and for the case $f=ggg$ we found perfect agreement after a couple of minor typographical errors were fixed in Ref.~\cite{Banerjee:2017faz}\footnote{We thank the authors of ~\cite{Banerjee:2017faz} for assistance with this comparison.}.

\subsection{Summary}
In this section we have presented the calculation of the hard function required to construct the $\tau_1 < \tau_1^{\text{cut}}$ part of our NNLO calculation. We have verified that our expressions reproduce the known IR limits at this order, and were able to reproduce a known result in the literature for decay kinematics.  Further we used our methodology to recompute the known processes for $H\rightarrow$ 3 partons. As a result of these checks we are therefore confident in using our results for the phenomenology presented in the subsequent sections of this paper. \\

\section{The $\tau_1 > \tau^{\text{cut}}_1$ contribution}
\label{sec:abovecut}

The remaining part of our calculation corresponds to the above cut piece where $\tau_1 > \tau_1^{\text{cut}}$. 
This piece corresponds to the NLO calculation of the process $A+2j$, allowing for a very soft jet requirement on the second jet to probe the regions with small $\tau_1$. 
The one-loop amplitude for the virtual part of the $A+2j$ process can be readily obtained from the literature by virtue of how the Higgs plus 4 parton one-loop amplitudes have been calculated.
The Higgs plus 4 parton one-loop amplitudes have been presented in a compact form several years ago~\cite{Dixon:2004za,Badger:2006us,Dixon:2009uk,Badger:2009hw,Badger:2009vh}. One of the reasons that the results 
are written so compactly is due to the amplitudes being calculated in terms of the following complex fields;
\begin{align}
\phi = \frac{H + i A}{2}, \qquad \phi^{\dagger} = \frac{H - i A}{2},
\end{align}
such that the amplitude for a Higgs plus parton amplitude is the sum of $\phi$ and $\phi^{\dagger}$ amplitudes 
\begin{align}
\mathcal{M}^{(1)} (H; \{p_k\}) 
=  
\mathcal{M}^{(1)} (\phi; \{p_k\}) + \mathcal{M}^{(1)} (\phi^{\dagger}; \{p_k\}).
\end{align}
On the other hand  we can generate pseudoscalar amplitudes from the difference of 
$\phi$ and $\phi^{\dagger}$ components,
\begin{align}
\mathcal{M}^{(1)} (A; \{p_k\}) 
= 
\frac{1}{i}
\left(
\mathcal{M}^{(1)} (\phi; \{p_k\}) - \mathcal{M}^{(1)} (\phi^{\dagger}; \{p_k\})
\right).
\end{align}
The simplicity arises from the observation that the $\phi$ field couples only to the self-dual part of the (trace of) the Gluon field strengths and the $\phi^{\dagger}$ couples only to the anti-self dual components. Due to this the helicity amplitudes for $\phi$ ($\phi^{\dagger}$) plus partons are considerably simpler than the corresponding $H$ or $A$ equivalents~\cite{Dixon:2004za}. 
By taking advantage of this fact, we have implemented the virtual part of the $A+2j$ process in MCFM 
with slight modifications of the scalar Higgs case. Specifically, for the NMHV helicity configurations in the literature 
only the $H$ amplitude has been defined and not the individual $\phi$ and $\phi^{\dagger}$ components. However (as a result of the helicity arguments discussed above) for an NMHV configuration 
only one of the amplitudes is complicated, the other being a simple rational function. We were therefore able to modify the results of the literature to extract 
the $\phi$ and $\phi^{\dagger}$ pieces separately and construct the $A+4$ parton amplitudes accordingly. 

The double-real part of our NNLO calculation requires the tree-level amplitudes for $A \to 5p$, which 
we have calculated using the BCFW recursion relations~\cite{Britto:2005fq}, obtaining compact analytic expressions. 
All the amplitudes present in the calculation have been checked against Madgraph~\cite{Alwall:2014hca} and GOSAM~\cite{GoSam:2014iqq} at random phase space points, finding excellent agreement.

The regularization of IR singularities present in the above-cut region has been performed using the dipole subtraction method~\cite{Catani:1996vz}. In the dipole method there are user-defined ``$\alpha$ parameters''~\cite{Nagy:1998bb} which 
determine the amount of (non-singular) phase space integrated over by the subtraction counter-terms.
An advantage of the method is that each type of dipole has a unique unphysical parameter which can be varied individually.
If the cancellation has proceeded correctly the individual real and virtual phase spaces depend on this choice of $\alpha$ but
the combined physical predictions do not.  Therefore following the procedure described in Ref.~\cite{Campbell:2019gmd} we define the following quantities 
\begin{equation}
\epsilon^{ab} = \frac{\sigma(\alpha_{ab} = 1) - \sigma(\alpha_{ab} = 0.01)}{\sigma(\alpha_{ab} = 1)} \,,
%\qquad
%a,b = I({\rm initial}), F({\rm final}) \,,
\label{eq:alpha_dep}
\end{equation}
where $a$ and $b$ are drawn from the set of initial and final $\{I,F\}$ type dipoles. We then proceed to calculate the cross sections using the following 
phase space requirements 
\begin{eqnarray}
{\rm LHC},~\sqrt{s} = 13~{\rm TeV}, \qquad && \mu_R = \mu_F = m_A = 125~{\rm GeV}, \nonumber \\
p_T^{\rm jet} > 20~{\rm GeV}, \qquad && \Delta R = 0.4 \nonumber \\
{\rm anti-}k_T, \qquad && \text{no explicit cut on rapidities} \,.
\end{eqnarray}
\begin{figure}
\begin{center}
\includegraphics[width=11cm]{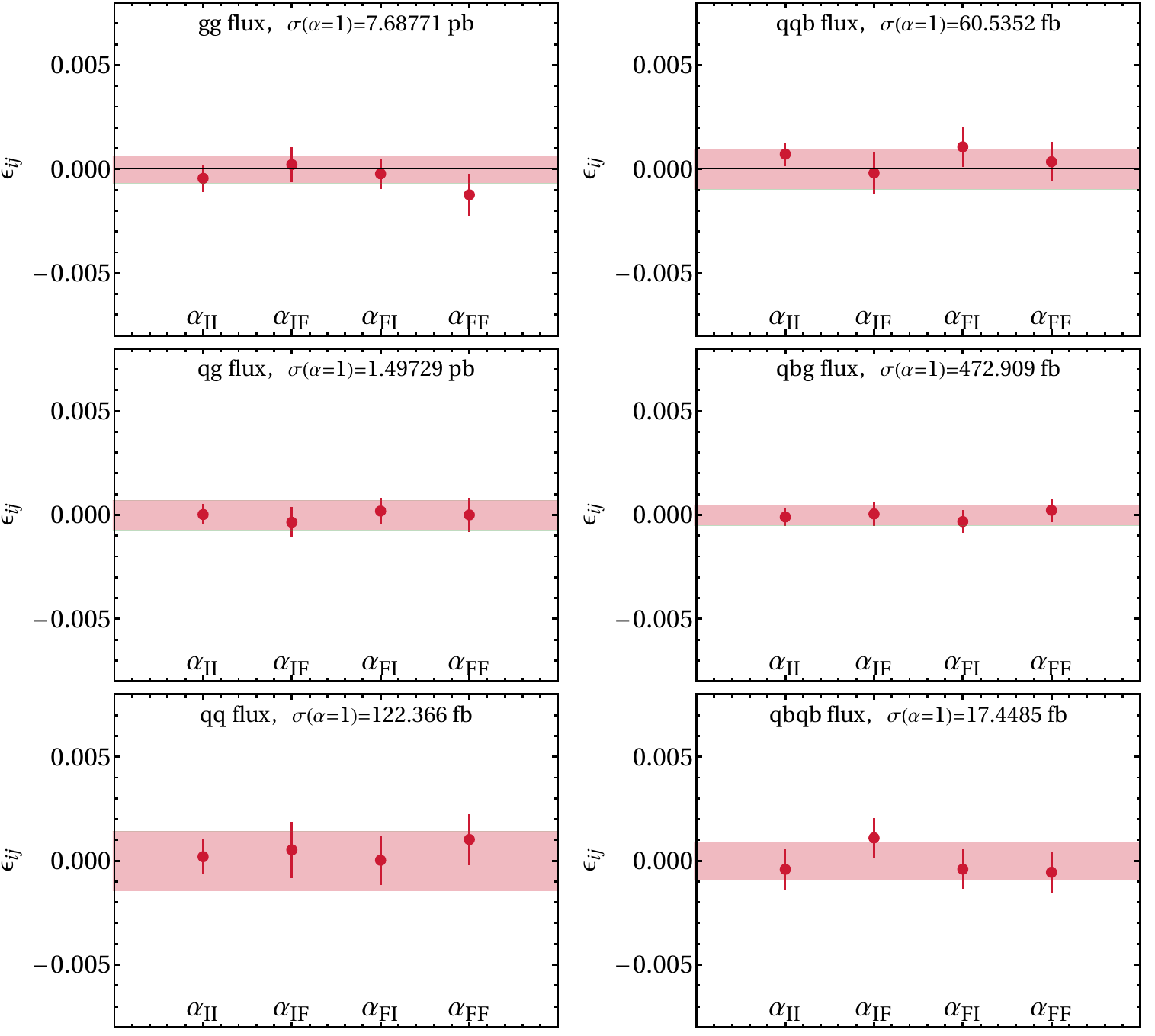}
\caption{The dependence of the $A+2j$ cross-section on the $\alpha$ parameters, for each partonic channel.
The points represent the deviation from $\alpha_{ab} = 1$ to $10^{-2}$.
The cross-sections obtained at default parameters ($\alpha_{ab} = 1$), are indicated in the plots.  
The horizontal lines represent the uncertainty on
a fit of the results to a constant. The plots show excellent $\alpha$-independence within the MC uncertainty range.}
\label{fig:alpha_dep} 
\end{center}
\end{figure}
Our results for the $\alpha$-(in)dependence are presented in Fig.~\ref{fig:alpha_dep}, where we have broken down cross sections into individual partonic sub-channels and investigated each type of dipole separately. In all cases excellent agreement is obtained with $\epsilon^{ab}=0$ within the (sub)per-mille level MC uncertainty.

In summary we have calculated all helicity amplitudes at one-loop level for $A+4$ partons and tree-level for $A +5$ partons and implemented the results into MCFM.
We have checked the validity of the individual amplitudes against Madgraph, and performed intricate tests of the dipole cancellation. We are therefore in position 
to put the above cut pieces together with the below cut pieces discussed in the previous section and test our full NNLO calculation. We do this in the next section.

\section{Results}
\label{sec:results}
The cross-section expanded to NNLO accuracy can be defined as follows,
\begin{align}
\sigma_{NLO} 
&= 
\sigma_{LO} + \delta \sigma_{NLO}\, , 
\\
\sigma_{NNLO} 
&=
\sigma_{LO} + \delta \sigma_{NLO} + \delta \sigma_{NNLO}\, . 
\end{align}
In the following section we will present and discuss results at NLO and NNLO. We have taken the calculations described in Sections~\ref{sec:hardfunction} and~\ref{sec:abovecut}, and implemented them into the parton level Monte Carlo code MCFM, making use of the existing implementation of the $H+j$ process~\cite{Campbell:2019gmd} where possible. 
In order to facilitate a comparison with the production of a scalar Higgs we introduce the following phase space selection requirements; 
\begin{eqnarray}
{\rm LHC},~\sqrt{s} = 13~{\rm TeV}, && \quad \mu_R = \mu_F = m_A = 125~{\rm GeV}\, , \nonumber \\
p_T^{\rm jet} > 30~{\rm GeV}, && \quad {\rm anti-}k_T~{\rm algorithm},~\Delta R = 0.4 \, ,  \\
{\rm PDF~set:} && \quad {\tt PDF4LHC15\_nnlo\_30} \, .
\end{eqnarray}
We note that these cuts match those used to study various NNLO calculations of $H+j$ at NNLO in Refs.~\cite{Campbell:2019gmd,Chen:2014gva,Chen:2016zka,Bizon:2018foh}. 
We will use these cuts throughout this section.

\subsection{Validation}

We begin by validating our calculation. Since the SCET-based factorization theorem for the below cut pieces neglects power suppressed terms, a natural check is to ensure that the Monte Carlo code can be run in a manner within the on-set of asymptotic behavior. 
We use the same definition for $\tau_1^{\text{cut}}$ as Ref.~\cite{Campbell:2019gmd} namely:
\begin{align}
\tau^{\text{cut}}_1 = \epsilon \times \sqrt{m_A^2 + \left( p_T^{j_1} \right)^2} \, . 
\label{eq:dyntau}
\end{align}
Typically we will draw values of $\epsilon$ from the range, 
$2.5 \times 10^{-5} \le \epsilon \le 5 \times 10^{-4}$.
It has been shown~\cite{Campbell:2019gmd} that if the $N$-jettiness variable is evaluated in the so-called boosted frame (corresponding to the rest frame of the Higgs-jet final state system) the resulting dependence on the unphysical parameter $\tau^{\text{cut}}_1$ is softened, and asymptotic behavior is reached sooner. Therefore in this paper we evaluate the 1-jettiness  in the boosted frame. 
\\
The parametric form of the leading power corrections are well known and have the following structure 
%In order to get the fitted values at NLO and NNLO, we have chosen the following forms of $\epsilon$ expansion as
\begin{align}
\sigma_{NLO}(\epsilon) 
&= 
\sigma_{NLO}^{0} + c_0 \epsilon \log (\epsilon) + \cdots \, ,
\label{eq:NLO_fit}
\\
\delta \sigma_{NNLO}(\epsilon) 
&=
\delta \sigma_{NNLO}^{0} + c_0 \epsilon \log^3 (\epsilon) + \cdots \, ,
\label{eq:NNLO_fit}
\end{align}
where $\epsilon$ is defined through Eq.~\eqref{eq:dyntau}.
\begin{figure}
\begin{center}
\includegraphics[width=11cm]{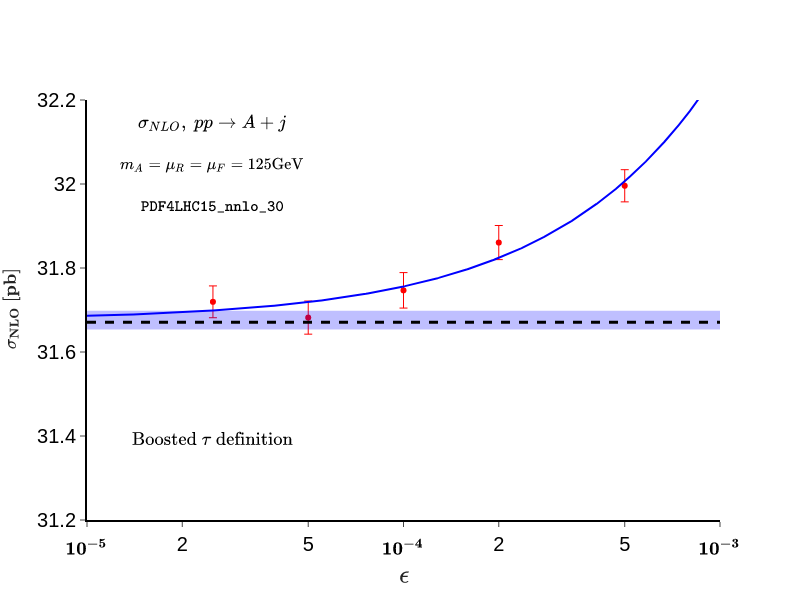}
\caption{$\tau$-dependence of the total NLO cross section, $\sigma_{NLO}$. 
The plot is made in the boosted frame. 
The blue solid line corresponds to the fitted curve from in Eq.~\eqref{eq:NLO_fit}, 
with the blue zone representing the errors from the fitted result.
The dipole subtraction is shown as the black dashed line.}
\label{fig:tau_dep_NLO} 
\end{center}
\end{figure}
In Fig.~\ref{fig:tau_dep_NLO} we present the results for the $A+j$ cross section at NLO using the default set of cuts. Shown in the figure is the result obtained using dipole subtraction (as the dashed line), the results from our calculation, and a fit to the results as described above. Using our parametric fit one can extract the following results 
\begin{align}
\sigma^{0}_{NLO} 
&= 31.674 \pm 0.022\ \text{pb} \,,
\\
\sigma^{dipole}_{NLO} 
&= 31.675 \pm 0.031\ \text{pb} \,.
\end{align}
which show the excellent agreement between the two methodologies at NLO accuracy. 
\begin{figure}
\begin{center}
\includegraphics[width=11cm]{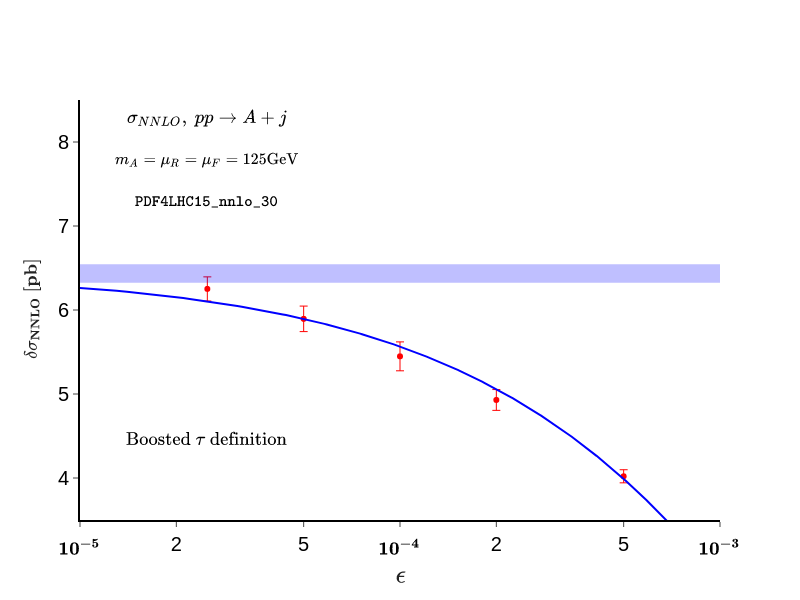}
\caption{
$\tau$-dependence of $\delta \sigma_{NNLO}$ using our default set of cuts. 
The blue solid line corresponds to the fit form in Eq.~\eqref{eq:NNLO_fit}, 
with the shaded blue color band representing the fitting errors.
}
\label{fig:tau_dep_NNLO} 
\end{center}
\end{figure}
We turn our attention to the NNLO coefficient in Fig.~\ref{fig:tau_dep_NNLO}. 
%describes the asymptotic behavior at NNLO nicely as well as the NLO case.
Again the results from our calculation are well modelled by the parametric fit (Eq.~\eqref{eq:NNLO_fit}). Performing the fit described above allows to extract the coefficient in the $\tau^{\text{cut}}_1 \rightarrow 0$ limit, obtaining: 
%The fitted result from the form of the Eq.~\eqref{eq:NNLO_fit} reads:
\begin{align}
\delta \sigma^{0}_{NNLO} 
&= 6.435 \pm 0.083\ \text{pb} \,.
\end{align}
\\
The general $\tau_1^{\rm{cut}}$ dependence of the NNLO coefficient is unsurprisingly almost identical to that reported in the calculation of $H+j$~\cite{Campbell:2019gmd}. While the asymptotic limit is harder to reach due to the more intricate NNLO phase space, as $\epsilon \sim 2-3 \times 10^{-5}$ the results agree with the limit within the reported uncertainties. We therefore take these values as within the asymptotic region and used $\epsilon = 2.5 \times 10^{-5}$ as our default value for the phenomenological studies in the next section. 
\\
\\
\subsection{Phenomenology}

In this section we explore the phenomenology of psuedoscalar Higgs plus jet production at NNLO. We keep the same fiducial selection criteria described in the previous section. NNLO predictions shown in this section are calculated using $\epsilon=2.5 \times 10^{-5}$ with the boosted $\tau$ definition. In our predictions we take $\mu_F=\mu_R=m_A$ as a central scale choice. We then perform a six-point scale variation  i.e. we compute distributions taking the extremeums from the set $(\mu_R/m_A , \mu_F/m_A )= (\alpha, \beta)$ where $(\alpha, \beta) \in \{(1,2),(1,1/2),(1/2,1),(2,1),(2,2),(1/2,1/2)\}$ and we exclude choices in which the scales are increased and decreased in opposite directions.

%Now we explore the scale dependence at NLO and NNLO for the rest of 
%this section.
%In order for it, in the following discussion, 
%we investigate $K$-factor of the cross sections across various 
%pseudoscalar Higgs masses, $p_T^{A}$ spectrum and $y^{A}$ distribution at $m_A = 125$ GeV via independent scale variation
%$1/2 m_A < \mu_R, \mu_F < 2 m_A$ with
%$1/2 < \mu_F / \mu_R < 2$. \\
\begin{figure}
\begin{center}
\includegraphics[width=11cm]{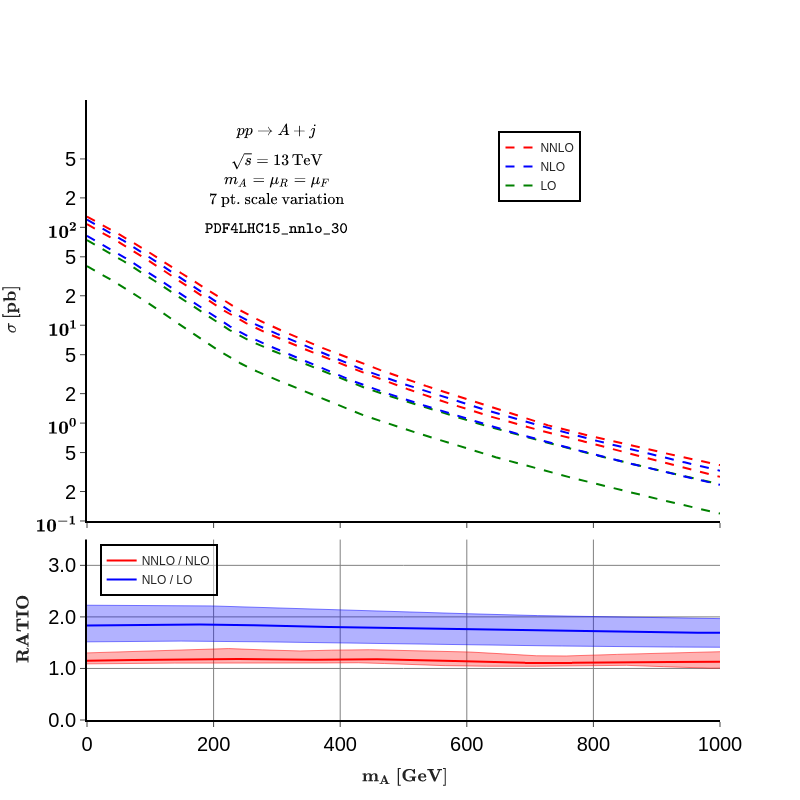}
\caption{Total cross sections and ratios of the cross sections across various pseudoscalar Higgs masses, using the our default cuts.
The error bands are obtained via variations of the renormalization and the factorization factor 
as described in the text.}
\label{fig:x_section} 
\end{center}
\end{figure}
In Fig.~\ref{fig:x_section} we present the total cross section for $A+j$ as a function of the mass of the pseudoscalar Higgs with $m_A$ taken in the range 0-1 TeV. As is expected from the analogous case in which the scalar Higgs is considered, the NNLO $K$-factor (the ratio of the NNLO to NLO cross sections) is sizable, around a factor of 1.2. The $K$-factor is reasonably flat across the mass range (with a gentle decrease as the mass increases). Going from NLO to NNLO significantly improves the scale variation, for instance at $m_A=125$ GeV the scale variation at NLO is 38 \%, which drops to 22 \% at NNLO. We also remind the reader that the PDF choice is the same across all predictions, larger differences are to be expected if lower order predictions matched to the relevant PDF set are used. It should be noted that in the region $m_A > 2m_t$ the effective field theory prescription becomes unreliable due to the resolved top quark. Significant work has been undertaken in the case of the scalar Higgs, and it has been shown that, if the cross section is weighted by the ratio of exact LO results including the mass, then the combined prediction does a good job of incorporating mass effects and higher order QCD corrections. Since a prediction resolving the top quark mass requires a (model dependent) definition of the $At\overline{t}$ vertex, we do not include such a re-weighting in this work (where we are primarily interested in the technical applications of the NNLO calculation). When performing a more detailed study of LHC phenomenology this is worth bearing in mind, but we leave this interesting study to future work. 
\\
\\
\begin{figure}
\begin{center}
\includegraphics[width=11cm]{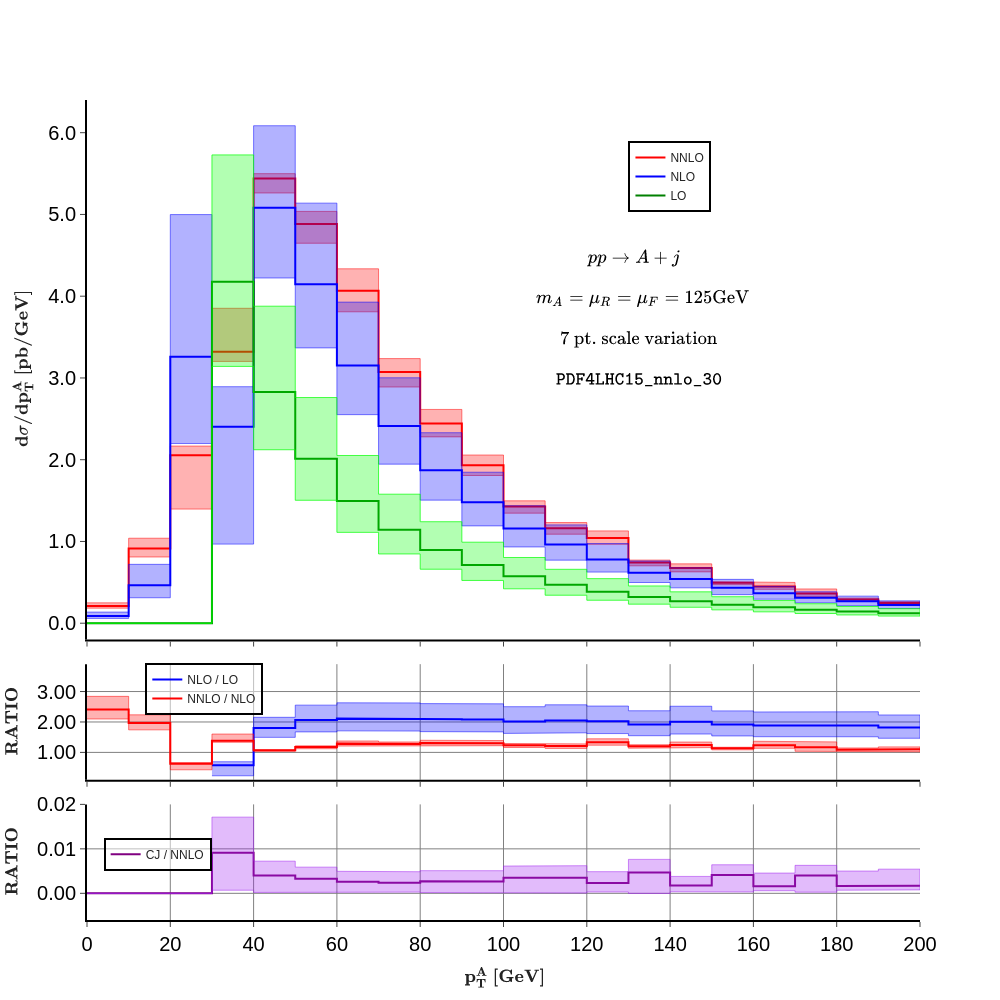}
\caption{$p_T^{A}$ distribution of the pseudoscalar Higgs ($m_A=125$ GeV) produced in association with an additional jet up to NNLO using our standard selection criteria.
The green, blue, and red colored histograms represent the LO, NLO, and NNLO predictions respectively.
The middle subplot shows $K$-factors which are the ratios of NLO over LO and NNLO over NLO, scale uncertainties are computed using a 7 point variation described in the text. For each ratio the scale choice in the denominator is fixed at $\mu_R=\mu_F=m_A$. The bottom panel shows the relative impact of the pieces proportional to the $C_J$ operator. 
\label{fig:pt_distribution_125}}
\end{center}
\end{figure}
\begin{figure}
\begin{center}
\includegraphics[width=11cm]{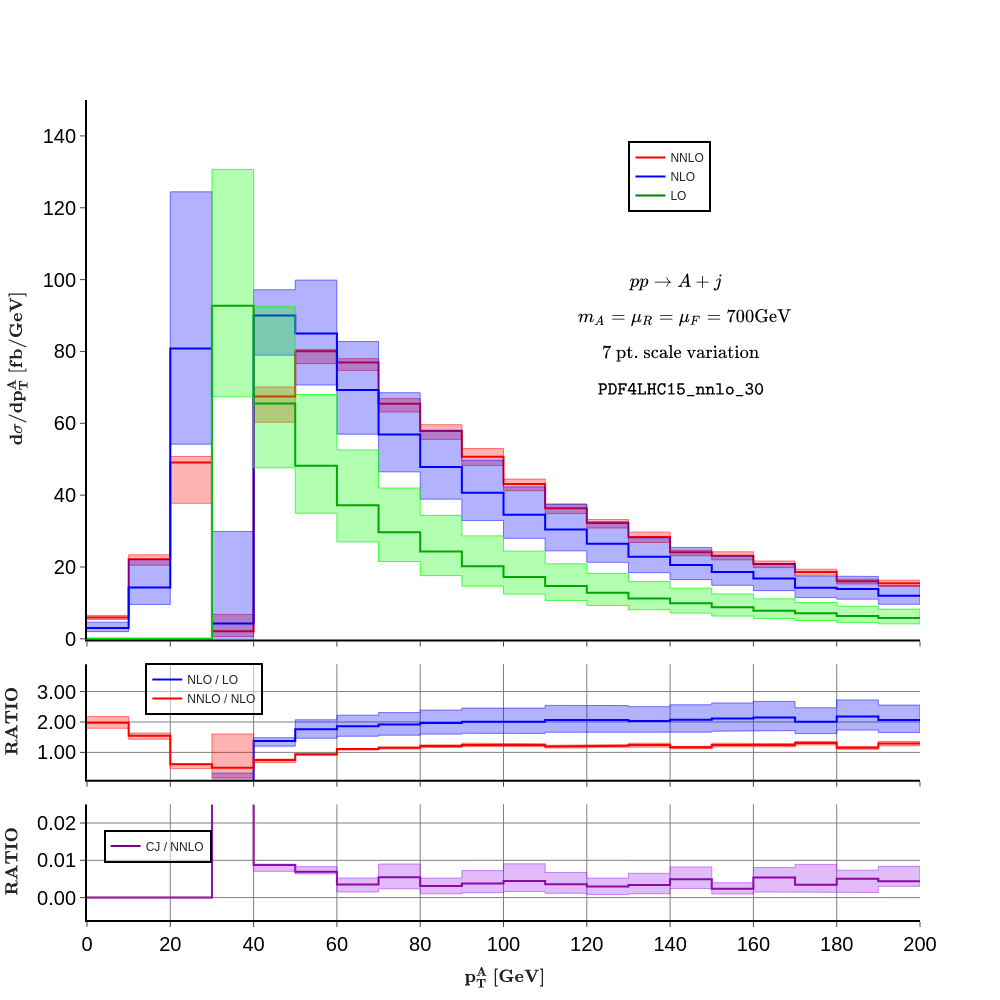}
\caption{$p_T^{A}$ distribution of the pseudoscalar Higgs evaluated with a mass of $m_A=700$ GeV. The remaining parameters are same as in Fig.~\ref{fig:pt_distribution_125}.
}
\label{fig:pt_distribution_700} 
\end{center}
\end{figure}
 Figs.~\ref{fig:pt_distribution_125}-\ref{fig:pt_distribution_700} show distribution of the transverse momentum of the pseudoscalar Higgs up to NNLO for two different mass choices $m_A=125$ and 700 GeV.  Including the NNLO corrections reduce the scale dependence compared the NLO prediction as expected. For instance, in the bulk phase space ($p_T^A >30$ GeV) the typical scale variation at NLO is around 44 \%, reduced to 13 \% at NNLO. 
We also note that the Sudakov shoulder effect~\cite{Catani:1997xc} is observed around $p_T^{A}$ = 30 GeV from the distribution as in the scalar Higgs case~\cite{Boughezal:2015aha}. In the region below $p_T^A < p_T^{j}$ only the real-virtual and real-real parts of the NNLO calculation contribute, so here the prediction is effectively NLO with the corresponding large $K$-factor and scale variation associated with predictions at this level. Finally in the lowest sub-plot we isolate the contributions sensitive to the $C_J$ operator. We recall that these pieces arise first at NNLO and contribute via an interference effect with the LO amplitude. As such, they reside in the $2\rightarrow 2$ phase space. These pieces are a small part of the total NNLO cross-section (around 0.5\% across the differential distribution), and they soften slightly at higher transverse momentum, which can be understood as an increasing relevance of the higher multiplicity phase spaces. 
\\
\\
\begin{figure}
\begin{center}
\includegraphics[width=11cm]{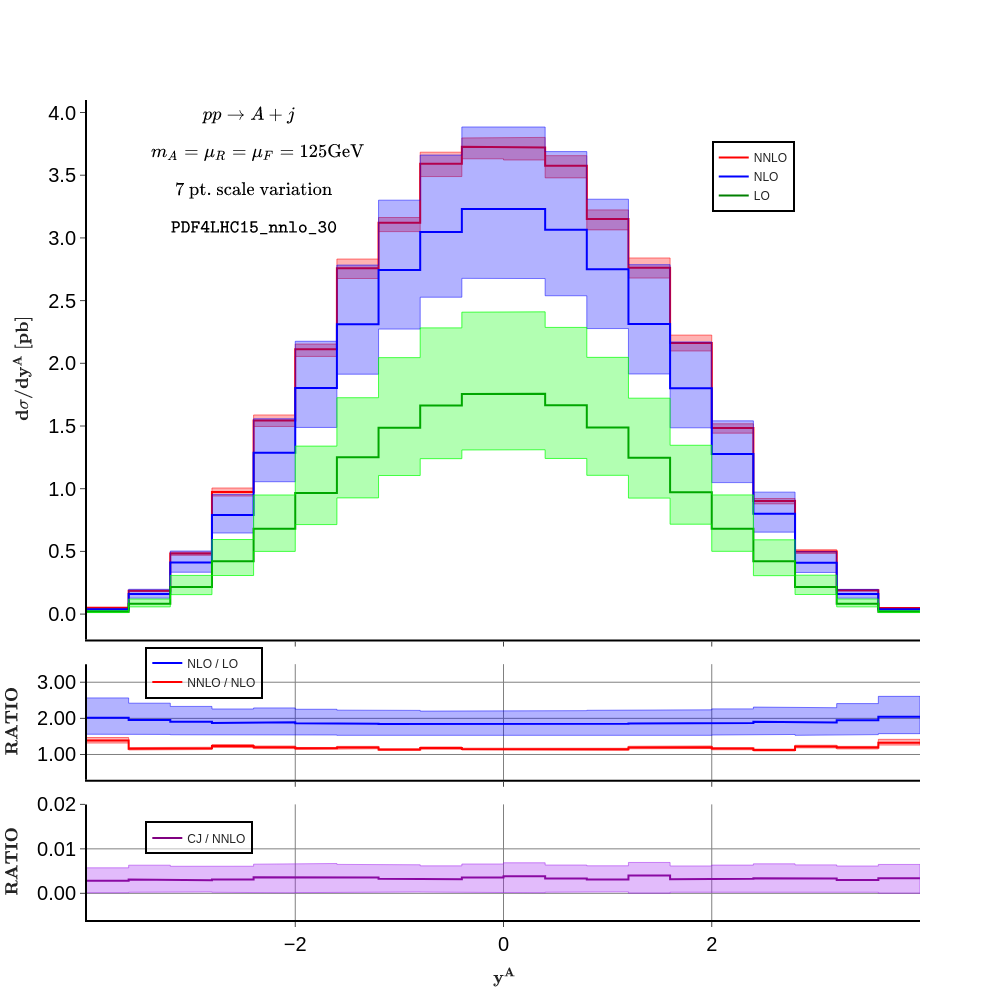}
\caption{The rapidity distribution of the pseudoscalar Higgs ($m_A=125$ GeV) produced in association with an additional jet up to NNLO.
The green, blue, and red colored histograms represent the LO, NLO, and NNLO respectively, scales are varied as described in the text. 
The middle subplot shows $K$-factors which are the ratios of NLO over LO and NNLO over NLO. The bottom subplot shows the relative impact of the $C_J$ operator. }
\label{fig:y_distribution_125} 
\end{center}
\end{figure}
\begin{figure}
\begin{center}
\includegraphics[width=11cm]{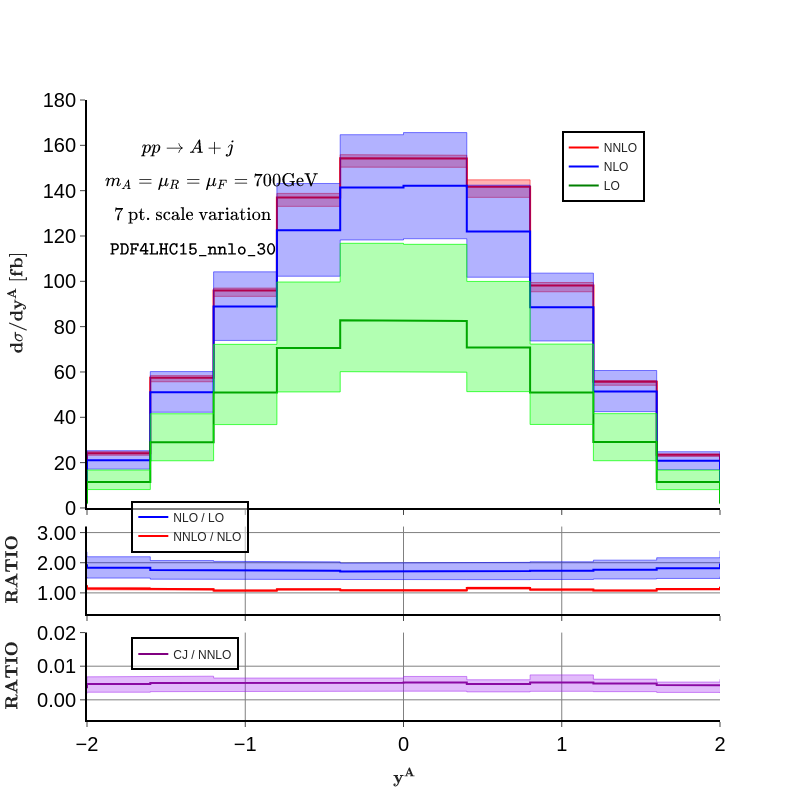}
\caption{The rapidity distribution of the pseudoscalar Higgs with $m_A$=700 GeV, all other parameters are the same as Fig.~\ref{fig:y_distribution_700}.
}
\label{fig:y_distribution_700} 
\end{center}
\end{figure}
Finally in Figs.~\ref{fig:y_distribution_125} and~\ref{fig:y_distribution_700} we present the rapidity distribution of the pseudoscalar Higgs at the same mass choices ($125$ and 700 GeV).  
%The histograms are normalized and 
The NNLO distribution is obtained in the same way as the $p^{A}_{T}$ distribution.
Whilst the effect the NLO and the NNLO corrections are approximately constant in rapidity as in the standard Higgs
case~\cite{Boughezal:2015aha,Campbell:2019gmd}, the NNLO correction 
significantly reduces the scale dependence as desired. Again the $C_J$ pieces represent a reasonably flat correction across the distribution, at around 0.5\% of the total NNLO prediction. 
\\
\\

%%%%%%%%%%%%%%%%%%%%%%%%%%%%%%%%%%%%%%%%%%%%%%%%%%%%
\section{Conclusions}
\label{sec:conclusion}

This paper presents the NNLO calculation of a psuedoscalar Higgs boson in association with an additional jet. The calculation of a psuedoscalar (or scalar) Higgs boson in association with a jet at this order is rather intricate, requiring a delicate combination of loop calculations and IR regulation. We have performed the calculation by independently re-calculating all components, including a recalculation of Higgs plus jet at this order to validate our psuedoscalar Higgs calculation at every stage. 
We calculated the two-loop amplitude for $A(H)\rightarrow$ 3 partons and the relevant crossings for LHC physics. We performed several checks of our calculation against both the literature and known IR limits of QCD amplitudes. 

In order to regulate the IR divergences present in the NNLO calculation we used the $N$-jettiness slicing method. Here one imposes a cut on the $1$-jettiness variable to separate the calculation into two regions. When the $1$-jettiness variable is sufficiently large the calculation has maximally one unresolved emission and resembles a traditional NLO calculation. Below the cut a factorization theorem from SCET allows the amplitude to be written in a convenient form (up to suppressed power corrections). 

We validated our calculation in several ways including; comparing to the literature results where appropriate, checking the dipole cancellation in the NLO $A+2j$ calculation and finally investigating the slicing dependence at NLO and NNLO to ensure asymptotic behaviour. We implemented the results of calculation into MCFM and used the code to produce phenomenological results at 13 TeV LHC. In particular we studied the transverse momentum and rapidity distribution of the pseudoscalar focusing on two representative masses. A natural extension of the work presented here is to include decays of the boson and investigate the impact on constraints on current searches at the LHC.

\acknowledgments  

We thank John Campbell, Prasanna Dhani, Thomas Gehrmann, Roberto Mondini  and  Ulrich Schubert for useful discussions. The authors are supported by the National Science Foundation through awards NSF-PHY-2014021 and NSF-PHY-2310363. Support provided by the Center for Computational Research at the University at Buffalo.

%%%%%%%%%%% appendix %%%%%%%%%%%%%%%%%%%%%%%%%%%%555
\appendix

\section{Formulae for renormalization and IR subtraction}
\label{renoirformulae}

The renormalization coefficients of Eqs.~\eqref{eq:Zalphas} and ~\eqref{eq:Z_GG}-\eqref{eq:Z_JJ} are defined as
\begin{align}
r_1 &= -\frac{\beta_0}{\epsilon} \,, \\
r_2 &= \frac{\beta_0^2}{\epsilon^2}-\frac{\beta_1}{2\epsilon} \,, \\
z_{GG,1} &= \frac{-11 C_A}{6 \epsilon} + \frac{N_F}{3\epsilon} \,, \\
z_{GG,2} &= \frac{1}{4\epsilon^2} 
\left( \frac{121 C_A^2}{9} - \frac{44 C_A N_F}{9} +\frac{4 N_F^2}{9} \right) 
+ \frac{1}{4\epsilon} \left( \frac{-17 C_A^2}{3} + \frac{5 C_A N_F}{3} + C_F N_F \right) \,, \\
z_{GJ,1} &= \frac{6 C_F}{\epsilon} \,, \\
z_{GJ,2} &= \frac{1}{2\epsilon^2} \left( -22 C_A C_F + 4 C_F N_F \right) 
+ \frac{1}{2\epsilon} \left( \frac{71 C_A C_F} {3} - 21 C_F^2 - \frac{2}{3}C_F N_F \right) \,, \\
z_{JJ,1} &= - 2 C_F \,, \\
z_{JJ,2} &= \frac{1}{4\epsilon} \left( \frac{22 C_A C_F}{3} + \frac{5 C_F N_F}{3} \right)
+ \frac{1}{4} \left(  22 C_F^2 - \frac{107 C_A C_F}{9} + \frac{31 C_F N_F}{18} \right) \,,
\end{align}
with
\begin{align}
\beta_0 &= \frac{11 C_A - 2 N_F}{6} \,, \\
\beta_1 &= \frac{17 C_A^2 - 10 C_A T_R N_F - 6 C_F T_R N_F}{6} \,,
\end{align}
and $T_R = \frac{1}{2}$, $C_A=N_c$, $C_F = \frac{N_c^2-1}{2 N_c}$. 

\vspace{5mm}

%%%%%%%%%%%%%%%%%%%%%%%%%%%%%%%%%%%%%%%%%%%%%%%%%%%%%%%%%%
The subtraction operators $\bI_f^{(\ell)}(\epsilon)$ for generic QCD processes can be found in Ref.~\cite{Catani:1998bh}. For completeness, we show here the explicit expressions for the subtraction operators in CDR.

For the process $A \rightarrow ggg, q\bar{q}g$ at the one-loop order:
\begin{align}
\bI_{ggg}^{(1)}(\epsilon) &= 
\frac{-e^{\epsilon \gamma_E}}{2 \Gamma(1-\epsilon)} \left (-\frac{m_H^2}{\mu^2}\right )^{-\epsilon}
\bigg[
N_C \left( \frac{1}{\epsilon^2} + \frac{\beta_0}{\epsilon N_C }
    \left( x^{-\epsilon} + y^{-\epsilon} + z^{-\epsilon} \right)
\right)
\bigg] \,,
\\
\bI_{q\bar{q}g}^{(1)}(\epsilon) &= 
\frac{-e^{\epsilon \gamma_E}}{2 \Gamma(1-\epsilon)} \left (-\frac{m_H^2}{\mu^2}\right )^{-\epsilon}
\bigg[
    N_C \left( \frac{1}{\epsilon^2} + \frac{3}{4 \epsilon} + \frac{\beta_0}{2\epsilon N_C} \right)
    \left( y^{-\epsilon} + z^{-\epsilon} \right)
    - \frac{1}{N_C} \left( \frac{1}{\epsilon^2} + \frac{3}{2\epsilon} \right) x^{-\epsilon}
\bigg] \,.
\end{align}
At the two-loop order for each $f=ggg, q\bar{q}g$:
\begin{align}
\bI_{f}^{(2)}(\epsilon) &=  
-\frac{1}{2} \bI_f^{(1)}(\epsilon) \bI_f^{(1)}(\epsilon)
-\frac{\beta_0}{\epsilon} \bI_f^{(1)}(\epsilon)
+e^{-\epsilon \gamma_E } \frac{ \Gamma(1-2\epsilon)}{\Gamma(1-\epsilon)} 
\left(\frac{\beta_0}{\epsilon} + K\right)
\bI_f^{(1)}(2\epsilon) + \boldsymbol{H}_f^{(2)}(\epsilon) \,,
\end{align}
where
\begin{align}
K &= \left ( \frac{67}{18} - \frac{\pi^2}{6} \right ) C_A - \frac{10}{9} T_R N_F \,, \\
\boldsymbol{H}_{ggg}^{(2)}(\eps) &= \frac{e^{\epsilon \gamma_E}}{4\epsilon \Gamma(1-\epsilon)} 
\left [ 3 H^{(2)}_g \right ] \,, \\
\boldsymbol{H}_{q\bar{q}g}^{(2)}(\eps) &= \frac{e^{\epsilon \gamma_E}}{4\epsilon \Gamma(1-\epsilon)} 
\left [ 2 H^{(2)}_q + H^{(2)}_g \right ] \,,
\end{align}
with
\begin{align}
H^{(2)}_g &= 
\left( \frac{\zeta_3}{2} + \frac{5}{12} + \frac{11 \pi^2}{144} \right) N_C^2
+ \frac{5}{27} N_F^2
+ \left( \frac{-\pi^2}{72} - \frac{89}{108} \right) N_C N_F 
- \frac{N_F}{4 N_C} \,,
\\
H^{(2)}_q &= 
\left( \frac{7 \zeta_3}{4} + \frac{409}{864} - \frac{11\pi^2}{96} \right) N_C^2 
+ \left( \frac{-\zeta_3}{4} - \frac{41}{108} - \frac{\pi^2}{96} \right) 
\notag \\
&+ \left( \frac{-3 \zeta_3}{2} - \frac{3}{32} + \frac{\pi^2}{8} \right) \frac{1}{N_C^2}
+ \left( \frac{\pi^2}{48} - \frac{25}{216} \right) \frac{(N_C^2 - 1)N_F}{N_C}
\, .
\end{align}

Lastly, we present the expressions for $\boldsymbol{Z}_f^{(1)}$ and $\boldsymbol{Z}_f^{(2)}$ in Eqs.~\eqref{mscoeff1} and \eqref{mscoeff2}. \\
For $f=q\bar{q}g$:
\begin{align}
\boldsymbol{Z}_{q\bar{q}g}^{(1)} &= 
\frac{1}{\epsilon^2} \left( N_C - \frac{1}{2 N_C} \right)
+ \frac{1}{\epsilon} \left[
\frac{N_C}{6} \left( -3 L(y)-3 L(z)+10 \right)
+ \frac{1}{4N_C} \left( 2 L(x)-3 \right)
- \frac{N_F}{6}
%\frac{-9 + 6L(x) +(20-6L(y)-6L(z))N_C^2 - 2N_C N_F}{12 N_C}
\right] \,,
\\
\boldsymbol{Z}_{q\bar{q}g}^{(2)} &= 
\frac{1}{\epsilon^4} \Bigg[
\frac{N_C^2}{2}+\frac{1}{8 N_C^2}-\frac{1}{2}
\Bigg] \notag \\
&+
\frac{1}{\epsilon^3} \Bigg[
\frac{N_C^2}{24} \bigg( -12 L(y)-12 L(z)+7 \bigg)
+ \frac{1}{N_C^2} \bigg( 3 - 2 L(x) \bigg) \notag \\
&+ \frac{1}{48} \bigg( 24 L(x)+12 L(y)+12 L(z)-43 \bigg)
+ \frac{N_C N_F}{12} - \frac{N_F}{24 N_C}
%-\frac{L(x)}{4 N_C^2}-\frac{1}{2} N_C^2 L(y)-\frac{1}{2} N_C^2 L(z)+\frac{N_F N_C}{12}-\frac{\text{Nf}}{24 N_C}+\frac{7 N_C^2}{24}+\frac{3}{8 N_C^2} \notag\\
%&+\frac{L(x)}{2}+\frac{L(y)}{4}+\frac{L(z)}{4}-\frac{43}{48}
\Bigg] \notag \\
&+
\frac{1}{\epsilon^2} \Bigg[
\frac{N_C^2}{24} \bigg( L(y) (6 L(z)-9)+3 L(y)^2+3 L(z)^2-9 L(z)-\pi ^2+19 \bigg) \notag\\
&+ \frac{1}{32 N_C^2} \bigg( 3 - 2 L(x) \bigg)^2 
+ \frac{N_C N_F}{72} + \frac{5 N_F}{72 N_C} - \frac{N_F^2}{72} \notag \\
&+ \frac{1}{144} \bigg( -18 L(x) (2 L(y)+2 L(z)-3)+54 L(y)+54 L(z)+3 \pi^2-148 \bigg)
%\frac{N_C N_F}{72}+\frac{5 N_F}{72 N_C}+\frac{L(x)^2}{8 N_C^2}-\frac{3 L(x)}{8 N_C^2}+\frac{1}{4} N_C^2 L(y) L(z)+\frac{1}{8} N_C^2 L(y)^2 \notag \\
%&-\frac{3}{8} N_C^2 L(y)+\frac{1}{8} N_C^2 L(z)^2-\frac{3}{8} N_C^2 L(z)-\frac{1}{24} \pi ^2 %N_C^2+\frac{19 N_C^2}{24}+\frac{9}{32 N_C^2}-\frac{N_F^2}{72} \notag \\
%&-\frac{1}{4} L(x) L(y)-\frac{1}{4} L(x) L(z)+\frac{3 L(x)}{8}+\frac{3 L(y)}{8}+\frac{3 L(z)}{8}+\frac{\pi ^2}{48}-\frac{37}{36}
\Bigg] \notag \\
&+
\frac{1}{\epsilon} \Bigg[
\frac{N_C^2}{1728} \bigg(
-1728 \zeta_3+24 \left(3 \pi ^2-67\right) L(y)+24 \left(3 \pi^2-67\right) L(z)+24 \pi ^2+4771
\bigg) \notag\\
&+\frac{1}{64 N_C^2} \bigg( 48 \zeta _3-4 \pi ^2+3 \bigg)
+ \frac{1}{864} \bigg(108 \zeta _3+\left(804-36 \pi ^2\right) L(x)+9 \pi ^2-1042\bigg) \notag\\
&+\frac{N_C N_F}{216} \bigg( 30 L(y)+30 L(z)-3 \pi ^2-110 \bigg)
+\frac{N_F}{432 N_C} \bigg(-60 L(x)+9 \pi ^2+92\bigg)
%-\frac{5 N_F L(x)}{36 N_C}+\frac{5}{36} N_C N_F L(y)+\frac{5}{36} N_C N_F L(z)-\frac{1}{72} \pi ^2 N_C N_F \notag \\
%&-\frac{55 N_C N_F}{108}+\frac{\pi ^2 N_F}{48 N_C}+\frac{23 N_F}{108 N_C}+\frac{1}{24} \pi ^2 N_C^2 L(y)-\frac{67}{72} N_C^2 L(y)+\frac{1}{24} \pi ^2 N_C^2 L(z) \notag \\
%&-\frac{67}{72} N_C^2 L(z)-\zeta _3 N_C^2+\frac{3 \zeta _3}{4 N_C^2}+\frac{1}{72} \pi ^2 N_C^2+\frac{4771 N_C^2}{1728}-\frac{\pi ^2}{16 N_C^2}+\frac{3}{64 N_C^2}+\frac{\zeta _3}{8} \notag \\
%&-\frac{1}{24} \pi ^2 L(x)+\frac{67 L(x)}{72}+\frac{\pi ^2}{96}-\frac{521}{432}
\Bigg] \,,
\end{align}
and for $f=ggg$:
\begin{align}
\boldsymbol{Z}_{ggg}^{(1)} &= 
\frac{1}{\epsilon^2} 
\left( 
\frac{3 N_C}{2}
\right)
+ \frac{1}{\epsilon} 
\left( 
\frac{N_C}{4} (-2 L(x)-2 L(y)-2 L(z)+11) - \frac{N_F}{2}
%-\frac{1}{2} N_C L(x)-\frac{1}{2} N_C L(y)-\frac{1}{2} N_C L(z)+\frac{11 N_C}{4}-\frac{N_F}{2}
\right) \,,
\\
\boldsymbol{Z}_{ggg}^{(2)} &= 
\frac{1}{\epsilon^4} \Bigg[
\frac{9 N_C^2}{8}
\Bigg] \notag \\
&+ \frac{1}{\epsilon^3} \Bigg[
\frac{-3 N_C^2}{16} \bigg( 4 L(x)+4 L(y)+4 L(z)-11 \bigg) - \frac{3}{8} N_C N_F
%-\frac{3 N_C N_F}{8}-\frac{1}{4} 3 N_C^2 L(x)-\frac{3}{4} N_C^2 L(y)-\frac{3}{4} N_C^2 L(z)+\frac{33 N_C^2}{16}
\Bigg] \notag \\
&+ \frac{1}{\epsilon^2} \Bigg[
\frac{N_C^2}{96} \bigg(8 L(x) (3 L(y)+3 L(z)-11)+12 L(x)^2+8 L(y) (3 L(z)-11) \notag\\
&+12 L(y)^2+12 L(z)^2-88 L(z)-6 \pi ^2+255\bigg) \notag\\
&+\frac{N_C N_F}{6} (L(x)+L(y)+L(z)-4) + \frac{N_F^2}{24}
%\frac{1}{6} N_C N_F L(x)+\frac{1}{6} N_C N_F L(y)+\frac{1}{6} N_C N_F L(z)-\frac{2 N_C N_F}{3}
%\notag \\
%&+\frac{1}{4} N_C^2 L(x) L(y)+\frac{1}{4} N_C^2 L(x) L(z)+\frac{1}{8} N_C^2 L(x)^2-\frac{11}{12} N_C^2 L(x)\notag \\
%&+\frac{1}{4} N_C^2 L(y) L(z)+\frac{1}{8} N_C^2 L(y)^2-\frac{11}{12} N_C^2 L(y)\notag \\
%&+\frac{1}{8} N_C^2 L(z)^2-\frac{11}{12} N_C^2 L(z)-\frac{1}{16} \pi ^2 N_C^2+\frac{85 N_C^2}{32}+\frac{N_F^2}{24}
\Bigg] \notag \\
&+ \frac{1}{\epsilon} \Bigg[
\frac{N_C^2}{288} \bigg(
-108 \zeta _3+4 \left(3 \pi ^2-67\right) L(x)+4 \left(3 \pi ^2-67\right) L(y) \notag\\
&+ 12 \pi ^2 L(z)-268 L(z)-33 \pi ^2+1384 \bigg) \notag\\
&+\frac{N_C N_F}{144} \bigg(20 L(x)+20 L(y)+20 L(z)+3 \pi ^2-155\bigg)
+ \frac{3 N_F}{16 N_C}
%\frac{5}{36} N_C N_F L(x)+\frac{5}{36} N_C N_F L(y)+\frac{5}{36} N_C N_F L(z)+\frac{1}{48} \pi ^2 N_C N_F-\frac{155 N_C N_F}{144}\notag \\
%&+\frac{3 N_F}{16 N_C}+\frac{1}{24} \pi ^2 N_C^2 L(x)-\frac{67}{72} N_C^2 L(x)+\frac{1}{24} \pi ^2 N_C^2 L(y)\notag \\
%&-\frac{67}{72} N_C^2 L(y)+\frac{1}{24} \pi ^2 N_C^2 L(z)-\frac{67}{72} N_C^2 L(z)-\frac{3}{8} \zeta _3 N_C^2-\frac{1}{96} 11 \pi ^2 N_C^2+\frac{173 N_C^2}{36}
\Bigg] \,,
\end{align}
where for brevity $L(a) = \text{ln} \! \left (-\frac{m_A^2}{\mu^2}\right ) + \text{ln} \,a$.

\section{Formulae for soft and collinear limits}
\label{matrixelements}
We list the unrenormalized $A \to gg$ ($\Lambda = G$) matrix elements that are needed for the collinear limit checks of the two-loop $A \to ggg, q\bar{q}g$ ($\Lambda = G$)  amplitudes. The matrix elements in CDR in which $D=4-2\epsilon$ and $\mu_R^2 = m_A^2$ read:
\begin{align}
\hat{\mathcal{M}}^{G}_{A \to gg} \hat{\mathcal{M}}^{G*}_{A \to gg}
&=
C_A C_F \left( C_G \right)^2 m_A^2
\Bigg[
\hat{\mathcal{M}}^{G,(0)}_{A \to gg} \hat{\mathcal{M}}^{G,(0)*}_{A \to gg}
+ \left( \frac{\alpha_s}{2\pi} \right) (-1)^{\epsilon} S_{\epsilon}
\hat{\mathcal{M}}^{G,(0)}_{A \to gg} \hat{\mathcal{M}}^{G,(1)*}_{A \to gg}
\notag \\
&+ \left( \frac{\alpha_s}{2\pi} \right)^2 (-1)^{2\epsilon} S_{\epsilon}^2
\hat{\mathcal{M}}^{G,(0)}_{A \to gg} \hat{\mathcal{M}}^{G,(2)*}_{A \to gg}
\Bigg] \,,
\end{align}
where
\begin{align}
\hat{\mathcal{M}}^{G,(0)}_{A \to gg} \hat{\mathcal{M}}^{G,(0)*}_{A \to gg} 
&= 1 -3 \epsilon + 2 \epsilon^2 \,,
\\
\hat{\mathcal{M}}^{G,(1)}_{A \to gg} \hat{\mathcal{M}}^{G,(0)*}_{A \to gg} 
&= 
N_C \Bigg[ 
\frac{-1}{\epsilon^2}
+ \frac{3}{\epsilon}
+ \frac{\pi ^2 }{12}
+ \epsilon \left( \frac{7 \zeta_3 }{3}-\frac{\pi ^2 }{4} \right)
+ \epsilon^2 \left( \frac{47 \pi ^4}{1440}-7 \zeta_3 \right)
\Bigg]
+ \mathcal{O}(\epsilon^3) \,,
\\
\hat{\mathcal{M}}^{G,(2)}_{A \to gg} \hat{\mathcal{M}}^{G,(0)*}_{A \to gg} 
&=
\frac{N_C^2}{2\epsilon^4}
+ \frac{1}{\epsilon^3} \Bigg[  \frac{N_C N_F}{12}-\frac{47 N_C^2}{24} \Bigg] 
\notag \\
&+ \frac{1}{\epsilon^2} \Bigg[  
%-\frac{N_C N_F}{9}+\frac{\zeta_2 N_C^2}{4}-\frac{1}{12} \pi ^2 N_C^2-\frac{5 N_C^2}{9} 
-\frac{N_C N_F}{9}+ N_C^2 \left( -\frac{5}{9}-\frac{\pi ^2}{24} \right)
\Bigg]
\notag \\
&+ \frac{1}{\epsilon} \Bigg[ 
%-\frac{1}{6} \zeta _2 N_C N_F-\frac{1}{72} \pi ^2 N_C N_F-\frac{161 N_C N_F}{108}+\frac{3 N_F}{4 N_C}+\frac{1}{6} \zeta _2 N_C^2 \notag\\
%&-\frac{25}{12} \zeta _3 N_C^2+\frac{47}{144} \pi ^2 N_C^2+\frac{811 N_C^2}{216}
N_C^2 \left( -\frac{25 \zeta _3}{12}+\frac{17 \pi ^2}{48}+\frac{811}{216} \right)
+ N_C N_F \left( -\frac{161}{108}-\frac{\pi ^2}{24} \right)
+ \frac{3 N_F}{4 N_C}
\Bigg]
\notag \\
&+ \Bigg[
%\frac{2}{9} \zeta _2 N_C N_F-\frac{19}{18} \zeta _3 N_C N_F-\frac{\zeta _3 N_F}{N_C}+\frac{1}{54} \pi ^2 N_C N_F-\frac{3965 N_C N_F}{648} 
%\notag \\
%&+\frac{71 N_F}{24 N_C}-\frac{19}{20} \zeta _2^2 N_C^2-\frac{1}{24} \pi ^2 \zeta _2 N_C^2-\frac{7}{18} \zeta _2 N_C^2+\frac{59}{9} \zeta _3 N_C^2+\frac{1}{240} \pi ^4 N_C^2
%\notag \\
%&+\frac{5}{54} \pi ^2 N_C^2+\frac{5941 N_C^2}{324}
N_C^2 \bigg( \frac{59 \zeta _3}{9}+\frac{\pi ^2}{36}-\frac{7 \pi ^4}{240}+\frac{5941}{324} \bigg)
+ N_C N_F \bigg( -\frac{19 \zeta _3}{18}+\frac{\pi ^2}{18}-\frac{3965}{648} \bigg)
\notag \\
&+ \frac{N_F}{N_C} \left( \frac{71}{24}-\zeta _3 \right)
\Bigg]
+ \mathcal{O}(\epsilon) \,,
\end{align}
which also can be found from Ref.~\cite{Ahmed:2015qpa} and $S_{\epsilon} = \frac{\exp{(\epsilon \gamma_E)}}{(4\pi)^{\epsilon}}$. The collinear functions in Eq.~\eqref{collinearlimiteq} are
for $f=ggg$,
\begin{align}
P_{ggg}^{(0)}(y,z) 
&=  \frac{ \left(z^2-z+1\right)^2}{(1-z) z} \,,
\\
P_{ggg}^{(1)}(y,z) 
&=
%\frac{1}{2 (1-z) z} \text{Sp}_{ggg,1}^{(1)}(\epsilon)
%+\left( \frac{z^3}{2 (1-z)}+\frac{(1-z)^3}{2 z} \right) \text{Sp}_{ggg,2}^{(1)}(\epsilon)
\frac{1}{(1-z) z} \text{Sp}_{ggg,1}^{(1)}(\epsilon)
+\left( \frac{z^3}{(1-z)}+\frac{(1-z)^3}{z} \right) \text{Sp}_{ggg,2}^{(1)}(\epsilon) \,,
\\
P_{ggg}^{(2)}(y,z) 
&=
%\frac{1}{2 (1-z) z} \text{Sp}_{ggg,1}^{(2)}(\epsilon)
%+\left( \frac{z^3}{2 (1-z)}+\frac{(1-z)^3}{2 z} \right) \text{Sp}_{ggg,2}^{(2)}(\epsilon)
\frac{1}{(1-z) z} \text{Sp}_{ggg,1}^{(2)}(\epsilon)
+\left( \frac{z^3}{(1-z)}+\frac{(1-z)^3}{z} \right) \text{Sp}_{ggg,2}^{(2)}(\epsilon)
\, ,
\end{align}
where $\text{Sp}_{ggg,n}^{(1)}(\epsilon)$ at $\mu_R^2 = m_A^2$ 
are defined as(see section 5. in~\cite{Badger:2004uk}.)
\begin{align}
\text{Sp}_{ggg,1}^{(1)}(\epsilon)
&=
(4 \pi)^{\epsilon} c_{\Gamma}(\epsilon)
(-y)^{-\epsilon}
\Bigg[
\frac{N_C}{2\epsilon^2}
\Bigg\{
-\Gamma(1-\epsilon) \Gamma(1+\epsilon) \left(\frac{z}{1-z}\right)^{\epsilon}
\notag \\
&+ \sum_{m=1}^{\infty} 2 \epsilon^{2m-1} \text{Li}_{2m-1} \left( \frac{1-z}{-z} \right)
\Bigg\}
+ \frac{z(1-z)}{(1-2\epsilon)(2-2\epsilon)(3-2\epsilon)} (N_C -N_C \epsilon - N_F)
\Bigg] \,,
\\
\text{Sp}_{ggg,2}^{(1)}(\epsilon)
&=
(4 \pi)^{\epsilon} c_{\Gamma}(\epsilon)
(-y)^{-\epsilon}
\Bigg[
\frac{N_C}{2\epsilon^2}
\Bigg\{
-\Gamma(1-\epsilon) \Gamma(1+\epsilon) \left(\frac{z}{1-z}\right)^{\epsilon}
\notag \\
&+ \sum_{m=1}^{\infty} 2 \epsilon^{2m-1} \text{Li}_{2m-1} \left( \frac{1-z}{-z} \right)
\Bigg\}
\Bigg] \,.
\end{align}
For $f=q\bar{q}g$,
\begin{align}
P_{q\bar{q}g}^{(0)}(x,z) 
&=
1-\frac{2 (1-z) z}{1-\epsilon} \,,
\\
P_{q\bar{q}g}^{(1)}(x,z) 
&= 
\left( 1-\frac{2 (1-z) z}{1-\epsilon} \right) \text{Sp}_{q\bar{q}g,1}^{(1)}(\epsilon) \,,
\\
P_{q\bar{q}g}^{(2)}(x,z) 
&= 
\left( 1-\frac{2 (1-z) z}{1-\epsilon} \right) \text{Sp}_{q\bar{q}g,1}^{(2)}(\epsilon)
\, .
\end{align}
where $\text{Sp}_{q\bar{q}g,1}^{(1)}(\epsilon)$ at $\mu_R^2 = m_A^2$
is defined as(see section 6. in~\cite{Badger:2004uk}.)
\begin{align}
\text{Sp}_{q\bar{q}g,1}^{(1)}(\epsilon)
&=
(4 \pi)^{\epsilon} c_{\Gamma}(\epsilon)
(-x)^{-\epsilon}
\Bigg[
\frac{N_C}{2\epsilon^2}
\Bigg\{
1-\Gamma(1-\epsilon) \Gamma(1+\epsilon) \left(\frac{z}{1-z}\right)^{\epsilon}
\notag \\
&+ \sum_{m=1}^{\infty} 2 \epsilon^{2m-1} \text{Li}_{2m-1} \left( \frac{1-z}{-z} \right)
\Bigg\}
\notag \\
&+ N_C \left(  \frac{13}{12\epsilon(1-2\epsilon)} + \frac{1}{6(1-2\epsilon)(3-2\epsilon)} \right)
+ \frac{1}{N_C} \left( \frac{1}{2\epsilon^2} + \frac{3}{4\epsilon(1-2\epsilon)} 
+ \frac{1}{2(1-2\epsilon)}\right)
\notag \\
& + N_F \left( \frac{-1}{3\epsilon(1-2\epsilon)} + \frac{1}{3(1-2\epsilon)(3-2\epsilon)} \right)
\Bigg] \, .
\end{align}
The two-loop splitting functions $\text{Sp}^{(2)}_{f,n}(\epsilon)$ in CDR are defined as:
\begin{align}
\text{Sp}^{(2)}_{f,n}(\epsilon) 
&= 
\frac{1}{2} \left ( \text{Sp}^{(1)}_{f,n}(\epsilon) \right )^2 
+ \frac{e^{-\epsilon \gamma_E}\,c_{\Gamma}(\epsilon)}{c_{\Gamma}(2\epsilon)} 
\left (\frac{\beta_0}{ \epsilon} + K \right ) \text{Sp}^{(1)}_{f,n}(2\epsilon) 
+ H_{f}(\epsilon) + \text{Sp}^{(2),\text{fin}}_{f,n} + \mathcal{O}(\epsilon) \, ,
\end{align}
where
\begin{align}
H_{ggg}(\epsilon) &= \frac{e^{-\epsilon \gamma_E}\,c_{\Gamma}(\epsilon)}{4\epsilon} 
(4\pi)^{2\eps} \, \left( -y \right )^{-2\epsilon} \left [ z (1-z) \right ]^{-2\epsilon} 
\left ( H^{(2)}_g - \beta_0 K + \beta_1 \right ) \, ,
\\
H_{q\bar{q}g}(\epsilon)
&=
\frac{e^{-\epsilon \gamma_E}\,c_{\Gamma}(\epsilon)}{4\epsilon} 
(4\pi)^{2\eps} \, \left( -x \right )^{-2\epsilon} \left [ z (1-z) \right ]^{-2\epsilon} 
\left ( 2 H^{(2)}_q - H^{(2)}_g - \beta_0 K + \beta_1 \right ) \, ,
\\
c_{\Gamma}(\epsilon) 
&= \frac{\Gamma(1-\epsilon)^2 \, \Gamma(1+\epsilon)}{\Gamma(1-2\epsilon)} \, .
\end{align}
with $H^{(2)}_g, H^{(2)}_q$ and $K$ as defined in Appendix \ref{renoirformulae}. Finally, the functions $\text{Sp}^{(2),\text{fin}}_{f,n}$ correspond to Eqs.~(5.18--5.20) and (6.8) of Ref.~\cite{Badger:2004uk} respectively with the replacement $w \to z$.

\section{IR-finite squared amplitudes and hard functions for $A \to ggg, q\bar{q}g$}
\label{app:NLO_results}
In this appendix we present the IR-finite squared amplitudes for $f=ggg, q\bar{q}g$ with $\mu_R^2 = m_A^2$\footnote{Note the results presented in Ref.~\cite{Banerjee:2017faz} are with the set $\mu_R^2 = -m_A^2$. }.
We normalized the amplitudes by Born color factor $2 C_A^2 C_F$ for $f=ggg$ and $C_A C_F$ for $q\bar{q}g$.
The results for the one-loop and two-loop hard functions in each kinematic region are also available in the attached auxiliary Mathematica files.
To save space, we only present up to the one-loop order for quick comparisons in this appendix. 
The higher order that the two-loop and the one-loop self interference amplitudes are provided in the attached auxiliary files.

\subsection{$f = ggg$ case}
The tree and one-loop results are given by
\begin{align}
\mathcal{M}_{ggg}^{G,(0),\text{fin}} \mathcal{M}_{ggg}^{G,(0),\text{fin}*}
&=
-2 C_{1,g} \, ,
\\
\text{Re}\left(\mathcal{M}_{ggg}^{G,(1),\text{fin}} \mathcal{M}_{ggg}^{G,(0),\text{fin}*}\right)
&=
N_C \Bigg[
C_{1,g} \bigg\{
H(0,y) H(0,z)-H(0,y) H(1,z)-\frac{11}{6} H(1-z,y) \notag \\
& -H(0,z) H(1-z,y)+2 H(1,z) H(z,y)-H(0,1-z,y) \notag \\
& -H(1-z,0,y)+2 H(z,1-z,y)+\frac{11}{6} H(0,y)+2 H(1,0,y) \notag \\
& +\frac{11}{6} H(0,z)-\frac{11}{6} H(1,z)+H(0,1,z)+H(1,0,z)+\frac{\pi ^2}{6}-4
\bigg\} \notag \\
& +\frac{C_{2,g}}{6}
\Bigg] 
+ N_F \Bigg[
\frac{C_{1,g}}{6} \bigg\{
2 H(1-z,y)-2 H(0,y)-2 H(0,z) \notag \\
& +2 H(1,z) 
\bigg\} - \frac{C_{2,g}}{6}
\Bigg] \, ,
\\
\text{Im}\left(\mathcal{M}_{ggg}^{G,(1),\text{fin}} \mathcal{M}_{ggg}^{G,(0),\text{fin}*}\right)
&=
C_{1,g} \left(
-\frac{11 \pi}{2}  N_C
\right)
+C_{1,g} N_F \pi \, ,
\\
\text{Re}\left(\mathcal{M}_{ggg}^{J,(1),\text{fin}} \mathcal{M}_{ggg}^{G,(0),\text{fin}*}\right)
&=
-2 C_{1,g} N_F \, ,
\\
\text{Im}\left(\mathcal{M}_{ggg}^{J,(1),\text{fin}} \mathcal{M}_{ggg}^{G,(0),\text{fin}*}\right)
&=0 \, ,
\end{align}
and the hard functions are given by
\begin{align}
\mathcal{M}_{ggg}^{G,(0),\text{ren}} \mathcal{M}_{ggg}^{G,(0),\text{ren}*}
&=
-2 C_{1,g} \, ,
\\
\text{Re}\left(\mathcal{M}_{ggg}^{G,(1),\text{ren}} \mathcal{M}_{ggg}^{G,(0),\text{ren}*}\right)
&=
N_C \Bigg[
C_{1,g} \bigg\{
H(0,y) H(0,z)-H(0,z) H(1-z,y) \notag \\
& -H(0,y) H(1,z)+H(1,z) H(1-z,y)+2 H(1,z) H(z,y) \notag \\
& -H(0,1-z,y)-H(1-z,0,y)+H(1-z,1-z,y) \notag \\
& +2 H(z,1-z,y)+H(0,0,y)+2 H(1,0,y)+H(0,0,z) \notag \\
& +H(0,1,z)+H(1,0,z)+H(1,1,z)
\bigg\}
- \frac{19 \pi^2}{12} C_{1,g} \notag \\
& -\frac{C_{3,g}}{6} \bigg\{ 24 y^4+48 y^3 z-48 y^3+72 y^2 z^2-143 y^2 z+71  y^2 \notag \\
& +48 y z^3-143 y z^2+142 y z-47 y+24 z^4-48 z^3+71 z^2 \notag \\
& -47 z+24
\bigg\}
\Bigg] - \frac{N_F}{6} C_{2,g} \, ,
\\
\text{Im}\left(\mathcal{M}_{ggg}^{G,(1),\text{ren}} \mathcal{M}_{ggg}^{G,(0),\text{ren}*}\right)
& =
C_{1,g} N_C  \bigg[ H(1-z,y)-H(0,y)-H(0,z)+H(1,z) \bigg] \pi \, ,
\\
\text{Re}\left(\mathcal{M}_{ggg}^{J,(1),\text{ren}} \mathcal{M}_{ggg}^{G,(0),\text{ren}*}\right)
& =
-2 C_{1,g} N_F \, ,
\\
\text{Im}\left(\mathcal{M}_{ggg}^{J,(1),\text{ren}} \mathcal{M}_{ggg}^{G,(0),\text{ren}*}\right)
& = 0 \, ,
\end{align}
where the constants $C_{i,g}$, $i=1,2,3$ are defined as
\begin{align}
C_{1, g} 
&= 
m_A^2 \frac{y^4+2 y^3 (z-1)+3 y^2 (z-1)^2+2 y (z-1)^3+\left(z^2-z+1\right)^2}{y z (y+z-1)} \, ,
\\
C_{2, g} 
&= 
m_A^2 \frac{(1-y) (1-z) (y+z)}{y z (-y-z+1)} \, ,
\\
C_{3, g} 
&=
\frac{m_A^2}{y z (y+z-1)} \, .
\end{align}
The amplitudes at NNLO (two-loop and one-loop$^2$) from the operator $O_G$ are decompose as following color terms:
\begin{align}
\mathcal{M}_{ggg}^{G,(2(1))} \mathcal{M}_{ggg}^{G,(0(1))*}
&=
N_C^2 \left( \mathcal{M}_{ggg}^{G,(2(1))} \mathcal{M}_{ggg}^{G,(0(1))*} \right)^{N_C^2}
+\frac{N_F}{N_C} \left( \mathcal{M}_{ggg}^{G,(2(1))} \mathcal{M}_{ggg}^{G,(0(1))*} \right)^{\frac{N_F}{N_C}}
\notag \\
&+N_F N_C \left( \mathcal{M}_{ggg}^{G,(2(1))} \mathcal{M}_{ggg}^{G,(0(1))*} \right)^{N_F N_C}
+N_F^2 \left( \mathcal{M}_{ggg}^{G,(2(1))} \mathcal{M}_{ggg}^{G,(0(1))*} \right)^{N_F^2}
\, .
\end{align}
Explicit expressions of the IR-finite amplitudes and the hard functions for each term can be found in the attached Mathematica files.

\subsection{$f = q\bar{q}g$ case}
The tree and one-loop results are given by
\begin{align}
\mathcal{M}_{q\bar{q}g}^{G,(0),\text{fin}} \mathcal{M}_{q\bar{q}g}^{G,(0),\text{fin}*}
&= - C_{1,q} \, ,
\\
\mathcal{M}_{q\bar{q}g}^{G,(0),\text{fin}} \mathcal{M}_{q\bar{q}g}^{J,(0),\text{fin}*}
&= 0 \, ,
\\
\mathcal{M}_{q\bar{q}g}^{J,(0),\text{fin}} \mathcal{M}_{q\bar{q}g}^{J,(0),\text{fin}*}
&= 0 \, ,
\\
\text{Re}\left(\mathcal{M}_{q\bar{q}g}^{G,(1),\text{fin}} \mathcal{M}_{q\bar{q}g}^{G,(0),\text{fin}*}\right)
& = N_C \Bigg[
\frac{C_{1,q}}{36} \bigg\{
-18 H(0,y) H(1,z)-18 H(0,z) H(1-z,y) \notag\\
& -39 H(1-z,y)+36 H(1,z) H(z,y) -18 H(0,1-z,y) \notag\\
& -18 H(1-z,0,y)+36 H(z,1-z,y)+30 H(0,y) \notag\\
& +18 H(1,0,y) +30 H(0,z)-39 H(1,z)+18 H(0,1,z)
\bigg\} \notag\\
& -\frac{C_{2,q}}{36} \bigg\{ 143 y^2-18 y z+9 y+143 z^2+9 z \bigg\}
\Bigg] \notag\\ 
& + \frac{1}{N_C} \Bigg[
\frac{C_{1,q}}{2} \bigg\{ -H(0,y) H(0,z)-H(1,0,y)-H(1,0,z) \bigg\} \notag\\
& - \frac{C_{2,q}}{12} \bigg\{ \pi ^2 \left(y^2+z^2\right)+21 y^2-6 y z+3 y+21 z^2+3 z \bigg\}
\Bigg] \notag\\ 
& + \frac{C_{1,q}}{36} N_F \Bigg[
12 H(1-z,y)-3 H(0,y)-3 H(0,z) \notag\\
& +12 H(1,z)+20
\Bigg] \, , 
\\
\text{Im}\left(\mathcal{M}_{q\bar{q}g}^{G,(1),\text{fin}} \mathcal{M}_{q\bar{q}g}^{G,(0),\text{fin}*}\right)
& = -\frac{11 \pi}{4}   C_{1,q} N_C 
+ \frac{\pi}{2} C_{1,q} N_F \, ,
\end{align}
\begin{flalign}
\text{Re}\left(
\mathcal{M}_{q\bar{q}g}^{G,(1),\text{fin}} \mathcal{M}_{q\bar{q}g}^{J,(0),\text{fin}*}
+\mathcal{M}_{q\bar{q}g}^{J,(1),\text{fin}} \mathcal{M}_{q\bar{q}g}^{G,(0),\text{fin}*}
\right)
& = - C_{1,q} N_F \, ,
\\
\text{Im}\left(
\mathcal{M}_{q\bar{q}g}^{G,(1),\text{fin}} \mathcal{M}_{q\bar{q}g}^{J,(0),\text{fin}*}
+\mathcal{M}_{q\bar{q}g}^{J,(1),\text{fin}} \mathcal{M}_{q\bar{q}g}^{G,(0),\text{fin}*}
\right)
& = 0 \, ,
&&
\end{flalign}
and the hard functions are given by
\begin{align}
\mathcal{M}_{q\bar{q}g}^{G,(0),\text{ren}} \mathcal{M}_{q\bar{q}g}^{G,(0),\text{ren}*}
&= - C_{1,q} \, ,
\\
\mathcal{M}_{q\bar{q}g}^{G,(0),\text{ren}} \mathcal{M}_{q\bar{q}g}^{J,(0),\text{ren}*}
&= 0 \, ,
\\
\mathcal{M}_{q\bar{q}g}^{J,(0),\text{ren}} \mathcal{M}_{q\bar{q}g}^{J,(0),\text{ren}*}
&= 0 \, ,
\\
\text{Re}\left(\mathcal{M}_{q\bar{q}g}^{G,(1),\text{ren}} \mathcal{M}_{q\bar{q}g}^{G,(0),\text{ren}*}\right)
& =  
N_C \frac{C_{1,q}}{12}  \Bigg[
-6 H(0,y) H(1,z)+12 H(1,z) H(z,y) \notag\\
& -6 H(0,z) H(1-z,y)-13 H(1-z,y)-6 H(0,1-z,y) \notag\\
& -6 H(1-z,0,y)+12 H(z,1-z,y)+6 H(0,0,y)+6 H(1,0,y) \notag\\
& -13 H(1,z)+6 H(0,0,z)+6 H(0,1,z)-7 \pi ^2-\frac{143}{3} + 3 C_{3,q}
\Bigg] \notag\\ 
& + \frac{1}{N_C} \frac{C_{1,q}}{4} \Bigg[
-2 H(0,y) H(0,z)-2 H(1,z) H(1-z,y) \notag\\
& -3 H(1-z,y)-2 H(1-z,1-z,y)-2 H(1,0,y)-3 H(1,z) \notag\\
& -2 H(1,0,z)-2 H(1,1,z)+\frac{5 \pi ^2}{6}-7 + C_{3,q}
\Bigg] \notag\\ 
& + N_F \frac{C_{1,q}}{9} \Bigg[
3 H(1-z,y)+3 H(1,z)+5
\Bigg] \, , 
\\
\text{Im}\left(\mathcal{M}_{q\bar{q}g}^{G,(1),\text{ren}} \mathcal{M}_{q\bar{q}g}^{G,(0),\text{ren}*}\right)
& = 
- \frac{C_{1,q}}{12} \pi N_C \bigg( 6 H(0,y)+6 H(0,z)+13 \bigg) \notag\\
& - \frac{C_{1,q}}{4} \pi \frac{1}{N_C} \bigg(  2 H(1-z,y)+2 H(1,z)+3 \bigg)
+ \frac{C_{1,q}}{3} \pi N_F \, ,
\end{align}
\begin{flalign}
\text{Re}\left(
\mathcal{M}_{q\bar{q}g}^{G,(1),\text{ren}} \mathcal{M}_{q\bar{q}g}^{J,(0),\text{ren}*}
+\mathcal{M}_{q\bar{q}g}^{J,(1),\text{ren}} \mathcal{M}_{q\bar{q}g}^{G,(0),\text{ren}*}
\right)
& = - C_{1,q} N_F \, ,
\\
\text{Im}\left(
\mathcal{M}_{q\bar{q}g}^{G,(1),\text{ren}} \mathcal{M}_{q\bar{q}g}^{J,(0),\text{ren}*}
+\mathcal{M}_{q\bar{q}g}^{J,(1),\text{ren}} \mathcal{M}_{q\bar{q}g}^{G,(0),\text{ren}*}
\right)
& = 0 \, ,
&&
\end{flalign}
where the constants $C_{i,q}$, $i=1,2,3$ are defined as
\begin{align}
C_{1, q} &= m_A^2 \frac{y^2+z^2}{y+z-1} \, , \\
C_{2, q} &= \frac{m_A^2}{y+z-1} \, , \\
C_{3, q} &= \frac{2 y z-y-z}{y^2+z^2} \, .
\end{align}
The amplitudes at NNLO (two-loop and one-loop$^2$) from the operator $O_G$ are decompose as following color terms:
\begin{align}
\mathcal{M}_{q\bar{q}g}^{G,(2(1))} \mathcal{M}_{q\bar{q}g}^{G,(0(1))*}
&=
N_C^2 \left( \mathcal{M}_{q\bar{q}g}^{G,(2(1))} \mathcal{M}_{q\bar{q}g}^{G,(0(1))*} \right)^{N_C^2}
+\frac{1}{N_C^2} \left( \mathcal{M}_{q\bar{q}g}^{G,(2(1))} \mathcal{M}_{q\bar{q}g}^{G,(0(1))*} \right)^{N_C^{-2}}
\notag \\
&+N_C^0 \left( \mathcal{M}_{q\bar{q}g}^{G,(2(1))} \mathcal{M}_{q\bar{q}g}^{G,(0(1))*} \right)^{N_C^0}
+\frac{N_F}{N_C} \left( \mathcal{M}_{q\bar{q}g}^{G,(2(1))} \mathcal{M}_{q\bar{q}g}^{G,(0(1))*} \right)^{\frac{N_F}{N_C}}
\notag \\
&+N_F N_C \left( \mathcal{M}_{q\bar{q}g}^{G,(2(1))} \mathcal{M}_{q\bar{q}g}^{G,(0(1))*} \right)^{N_F N_C}
+N_F^2 \left( \mathcal{M}_{q\bar{q}g}^{G,(2(1))} \mathcal{M}_{q\bar{q}g}^{G,(0(1))*} \right)^{N_F^2}
\, .
\end{align}
Explicit expressions of the IR-finite amplitudes and the hard functions for each term can be found in the attached Mathematica files.

\section{Recalculation of $H \to ggg, q\bar{q}g$ Amplitudes}
\label{sec:Higgs_recalculation}

The NNLO QCD corrections to $H+j$ have been well studied in the literature, and serve as a 
useful cross check of our calculation. Therefore, we have checked our results for $A+j$ by reproducing the relevant 
results for $H+j$ at every stage of our calculation. 
The most intricate cross check regards the reproduction of the two-loop amplitudes for $H \to ggg, q\bar{q}g$, which were originally presented in Ref.~\cite{Gehrmann:2011aa} and more recently expanded to higher orders in $\epsilon$, in Ref.~\cite{Gehrmann:2023etk}.
%While we could match most of results, however, we found out there is a small discrepancy from the analytic continued results. In this appendix, we describe it and represent our results. \\
In Refs.~\cite{Gehrmann:2011aa,Gehrmann:2023etk}, the helicity amplitudes $\alpha,\, \beta,$ and $\gamma$ are calculated
for the $H \to ggg, q\bar{q}g$ process up to two-loop order.
These are expanded to second order in $\alpha_s$ in terms of $\Omega = \alpha,\, \beta,\, \gamma$ as follows
\begin{align}
\Omega = C_H \sqrt{4 \pi \alpha_s} \, T_{\Omega}
\left[
\Omega^{(0)}
+ \left( \frac{\alpha_s}{2\pi} \right) \Omega^{(1)}
+ \left( \frac{\alpha_s}{2\pi} \right)^2 \Omega^{(2)}
+ \mathcal{O} (\alpha_s^3)
\right]\,,
\end{align}
where $C_H$ is the effective $Hgg$ coupling~\cite{Chetyrkin:1997iv,Chetyrkin:1997un} in the $m_t \to \infty$ limit and 
the color factor is $T_{\alpha} = T_{\beta} = f^{a_1 a_2 a_3}$ and $T_{\gamma} = T^{a_3}_{i_1 j_2}$.
We have been able to reproduce the same IR-finite results for  $\alpha,\, \beta,$ and $\gamma$
 as presented in appendix A and B in ref.~\cite{Gehrmann:2011aa}.
%From thos all necessary squared amplitudes in the CDR scheme can be assembled.
Since we do not use a projector-based approach for our $A+j$ calculation a second useful check it so calculate the squared amplitudes directly.
The squared amplitudes are expanded in order of $\alpha_s$ as
\begin{align}
\label{eq:Higgs_expand}
\mathcal{A}_{H \to f} \mathcal{A}_{H \to f}^* 
&= (4 \pi \alpha_s) \left( \frac{\alpha_s}{2 \pi} \right)^2 \left( C_H \right)^2 \mathcal{C}_f
\Bigg[  
\mathcal{M}_{H \to f}^{(0)} \mathcal{M}_{H \to f}^{(0)*} 
+ \left( \frac{\alpha_s}{2 \pi} \right) 
2 \,\text{Re}\left(  \mathcal{M}_{H \to f}^{(1)} \mathcal{M}_{H \to f}^{(0)*}  \right) \notag \\
&+ \left( \frac{\alpha_s}{2 \pi} \right)^2 
\bigg\{
2 \,\text{Re}\left(  \mathcal{M}_{H \to f}^{(2)} \mathcal{M}_{H \to f}^{(0)*} \right) 
+ \mathcal{M}_{H \to f}^{(1)} \mathcal{M}_{H \to f}^{(1)*} 
\bigg\}
\Bigg] \,,
\end{align}
where the Born color factor $\mathcal{C}_f$ is $2C_A^2 C_F$ and $C_A C_F$ 
for $f = ggg,\, q\bar{q}g$.
We have made use of the same program as described in section~\ref{sec:hardfunction} for calculating those and could reproduce the same IR-finite results as the squared amplitudes constructed from 
the $\alpha,\, \beta,$ and $\gamma$, which gives us confidence in the related calculation for the pseudo Higgs amplitudes. 
Furthermore, we have performed a test for soft and collinear limit behavior of the helicity amplitudes and they all are consistent with the results from Ref.~\cite{Badger:2004uk} in which the splitting functions have been extracted from Ref.~\cite{Gehrmann:2011aa}. %which gives us one more layer of %robustness of our calculation.\\
\\
In order to utilize the amplitudes described above for LHC scattering the partons must be crossed to the initial state via analytic continuation. 
We confirmed our crossed integrals numerically with AMFlow~\cite{Liu:2022chg} finding perfect agreement. We have also checked our crossed results against the literature and, found perfect agreement with the results from the recent paper~\cite{Gehrmann:2023etk}.  
We note that there is a small typo in Ref.~\cite{Gehrmann:2011aa} for $\gamma^{(2)}$ of the $q \bar{q} \to H g$ kinematic region (the region$(2a_+)$ as denoted in~\cite{Gehrmann:2002zr}). 
%\begin{table}
%\centering
%\begin{tabular}{|c|c|c|c|}
%\hline
%\rule{0pt}{2ex}{Color term from $\gamma^{(2)}_{q\bar{q} \to H %g}$} & Ref.~\cite{Gehrmann:2011aa} & Our result\\
%\hline
%$N_C^2$ & $-48.8936549+ 4.9993577 i$ & $-47.0845075+4.9993577 i$\\
%\hline
%$N_C^{-2}$ & $7.74370309+6.09760127 i$ & $4.12540814+6.09760127 %i$\\
%\hline
%$N_C^0 N_F^0$ & $1.91193443-13.31480035 i$ & $0.10278695-%13.31480035 i$\\
%\hline
%$N_F/N_C$ & $1.39742590+2.80487637 i$ & $1.39742590+2.80487637 %i$\\
%\hline
%$N_F N_C$ & $14.5429926-16.6731338 i$ & $14.5429926-16.6731338 %i$\\
%\hline
%$N_F^2$ & $-0.78937783+2.72921019 i$ & $-0.78937783+2.72921019 %i$\\
%\hline
%\end{tabular}
%\caption{Numerical comparison of $\gamma^{(2)}_{q \bar{q} \to H %g}$ for each color term. 
%The all inputs are same as Eq.~\eqref{eq:HiggsHardInput} but %$\mu_R = m_H = 0.1$ TeV.}
%\label{tab3}
%\end{table}
%To illustrate the difference we present the numerical evaluation %for each color structure in table \ref{tab3}. 
%The table shows that, aside from the $N_F$ terms there is a %disagreement in the real part of each color structure. 
%We also have compared singular behaviors in the collinear limit, however, the both show very similar results, and therefore, they are not distinguishable.
%We are not arguing our calculation is correct nor the Ref.~\cite{Gehrmann:2011aa} is incorrect.
%\\
Since the results of Ref.~\cite{Gehrmann:2011aa} have been widely used, this discrepancy can give small modifications to some other published articles. For example, in Ref.~\cite{Becher:2013vva}, the authors assembled the two-loop hard functions for $H+j$.
%for $H \to ggg, q\bar{q}g$
%by taking the helicity amplitudes from~\cite{Gehrmann:2011aa}. 
The numerical results for the hard functions are provided in Eq.(73) in section.5, 
which we present again here for the reader:
\begin{align}
\label{eq:Higgs_hards_numerical}
\hat{H}_{gg\to H g}(\hat{u},\hat{t},\mu) 
&= 1+  (6.02164) \,  \alpha _s +  (24.2724)  \,  \alpha _s^2\,, 
\notag\\
\hat{H}_{q \bar{q}\to H g}(\hat{u},\hat{t},\mu) 
&= 1+ (1.85023) \,\alpha _s +  (8.15565) \, \alpha _s^2\,, 
\notag\\
%\hat{H}_{q g\to H q}(\hat{u},\hat{t},\mu) &= 1+2.77875\, \alpha _s -  12.0751 \,\alpha _s^2 \,.
\hat{H}_{q g\to H q}(\hat{u},\hat{t},\mu) 
&=1 +  (6.63865) \, \alpha _s + (24.9851)  \,\alpha _s^2 \,,
\end{align}
at the following inputs:
\begin{align}
\label{eq:HiggsHardInput}
(\hat{s},\hat{t}) 
&= 
(1\,\text{TeV}^2,\, -0.4\,\text{TeV}^2) \,,
\notag \\
m_H &= 0.1 \,\text{TeV} \,,
\notag \\
\mu_R &= 0.6 \,\text{TeV} \,.
\end{align}
With our calculation we have reproduced the same numerical results for 
$\hat{H}_{gg\to H g}$ and $\hat{H}_{q g\to H q}\,$, but have obtained different
results for $\hat{H}_{q \bar{q}\to H g}$, namely, 
:
\begin{align}
\hat{H}_{q \bar{q}\to H g}(\hat{u},\hat{t},\mu)  &= 
1 + (1.85023) \,\alpha _s +  (7.54245) \, \alpha _s^2\,.
\label{eq:Hard_inputs}
\end{align}
\\
%Finally we note that, in Ref.~\cite{Campbell:2019gmd} the authors have taken the hard functions from
%Ref.~\cite{Becher:2013vva} for the MCFM implementation of the H+jet@NNLO process. 
A natural concern is that this difference may result in a phenomenological difference for the various calculations of 
$H+j$@NNLO, which used this hard function. 
%One can be concerned that the result from the Ref.~\cite{Campbell:2019gmd} can be affected, though,
However, due to the relative smallness of the $q\overline{q}$ initial state (as can be seen from the
breakdown of the partonic channels for instance in Ref.~\cite{Campbell:2019gmd}), the 
impact of the different hard function is negligible at the level of the total cross section. 
%We have not investigated the impact of the difference integrating out the final state gluon (which would be relevant for applications 
%to differential $H$@N3LO), here the difference may be more important. 
%we observed that an impact on the total cross section due to this discrepancy from the hard functions is marginal enough to be ignored.
We have assembled all the IR-finite squared amplitudes and the hard functions for each kinematic region up to NNLO for an arbitrary renormalization scale. These are attached as auxiliary files to the {\tt{arXiv}} submission, which may be useful in further comparisons of $H@$N3LO and $H+j$@NNLO.

\section{Descriptions of the ancillary files}
\label{app:ancillary_files}
In this appendix we describe the structure and the notation of the attached Mathematica ancillary files.
The IR-finite squared amplitudes for the SM Higgs are written in the files with a name which begins with
\texttt{Hggg_IR} and \texttt{Hqqg_IR} for the process $f=ggg$ and $f=q\bar{q}g$ respectively.
Likewise, the hard functions for the SM Higgs are presented in the files with a name which begins  
\texttt{Hggg_Hard} and \texttt{Hqqg_Hard}.
For the pseudoscalar Higgs case, the files are named in a similar manner which begins with 
\texttt{A} instead of \texttt{H}.
We have normalized the each amplitude for $f=ggg, q\bar{q}g$ processes by 
$2 C_A^2 C_F$ and $C_A C_F$  respectively.
Then any one-loop and two-loop(one-loop self interference) amplitudes can be decomposed as
\begin{align}
\mathcal{M}_{f}^{(1)} \mathcal{M}_{f}^{(0)*}
&=
N_C \left( \mathcal{M}_{f}^{(1)} \mathcal{M}_{f}^{(0)*} \right)^{N_C}
+
\frac{1}{N_C} \left( \mathcal{M}_{f}^{(1)} \mathcal{M}_{f}^{(0)*} \right)^{N_C^{-1}}
+
N_F \left( \mathcal{M}_{f}^{(1)} \mathcal{M}_{f}^{(0)*} \right)^{N_F}\,,
\\
\mathcal{M}_{f}^{(2(1))} \mathcal{M}_{f}^{(0(1))*}
&=
N_C^2 \left( \mathcal{M}_{f}^{(2(1))} \mathcal{M}_{f}^{(0(1))*} \right)^{N_C^2}
+
N_C^0 \left( \mathcal{M}_{f}^{(2(1))} \mathcal{M}_{f}^{(0(1))*} \right)^{N_C^0}
\notag \\
&+
\frac{1}{N_C^2} \left( \mathcal{M}_{f}^{(2(1))} \mathcal{M}_{f}^{(0(1))*} \right)^{N_C^{-2}}
+
\frac{N_F}{N_C} \left( \mathcal{M}_{f}^{(2(1))} \mathcal{M}_{f}^{(0(1))*} \right)^{\frac{N_F}{N_C}}
\notag \\
&+N_F N_C \left( \mathcal{M}_{f}^{(2(1))} \mathcal{M}_{f}^{(0(1))*} \right)^{N_F N_C}
+N_F^2 \left( \mathcal{M}_{f}^{(2(1))} \mathcal{M}_{f}^{(0(1))*} \right)^{N_F^2}
\,.
\end{align}
The evolution of the renormalization scale $\mu_R$ has been carried for the all orders
in the attached files.
For the LHC application the squared amplitudes need to be crossed to the appropriate
kinematic regions. 
To do so, the required analytic continuations are from 
the region(1) to the region(2) and (4)(according to the Ref.~\cite{Becher:2013vva, Gehrmann:2002zr}) such as
\begin{align}
g(p_1) + g(p_2) \to A(H)(p_4) + g(-p_3)
\end{align}
from $f=ggg$ and
\begin{align}
q(p_1) + \bar{q}(p_2) \to A(H)(p_4) + g(-p_3)\,, \quad
\bar{q}(p_2) + g(p_3) \to A(H)(p_4) + \bar{q}(-p_1)
\end{align}
from $f=q\bar{q}g$.
The dimensionless variables for the region(2) and the region(4)
are defined as~\cite{Becher:2013vva, Gehrmann:2002zr}:
\begin{align}
u_2 = -\frac{y}{x}\,, \qquad v_2 = \frac{1}{x}\,,
\\
u_4 = -\frac{y}{z}\,, \qquad v_4 = \frac{1}{z}\,.
\end{align}
The IR-finite squared amplitudes and the hard functions for the each region is attached to the file name ends with 
\texttt{R1, R2}, and \texttt{R4} respectively.
The notation of the attached Mathematica files are as follows:
\begin{align}
N_C &\to \texttt{Nc}\,, \qquad  N_F \to \texttt{Nf},
\notag \\
u_{2,4} &\to \texttt{u}\,, \qquad  v_{2,4} \to \texttt{v},
\notag \\
\zeta_2, \zeta_3, \zeta_4  &\to \texttt{zeta2, zeta3, zeta4}\,,   
\notag \\
\frac{\mu_R^2}{m_H^2} &\to \texttt{muSqOMH2}\,, \quad  \frac{\mu_R^2}{m_A^2} \to \texttt{muSqOMA2} \,.
\end{align}
For the convention of the $1D$ and $2D$ HPLs we followed the notation introduced in the Ref.~\cite{Gehrmann:2000zt}.
For instance, we denote
\begin{align}
H(0,1,1,0;x) \to \texttt{H[0,1,1,0,x]} \,, \qquad  H(1,z,1-z;y) \to \texttt{H[1,z,1-z,y]}\,.
\end{align}

The required terms at $\text{LO}(\mathcal{O}(\alpha_s^3))$, $\text{NLO}(\mathcal{O}(\alpha_s^4))$, 
and $\text{NNLO}(\mathcal{O}(\alpha_s^5))$ 
expanded as~\eqref{eq:Higgs_expand}, ~\eqref{eq:ggg_amp_sq}, and ~\eqref{eq:qqg_amp_sq} 
for the Higgs and the pseudoscalar Higgs are written in the files for the each kinematic region.
The notation for the terms for the Higgs case are written as:
\begin{align}
\mathcal{M}_{H \to f}^{(0)} \mathcal{M}_{H \to f}^{(0)*} 
&\to 
\texttt{M0M0}\,, 
\qquad
\mathcal{M}_{H \to f}^{(1)} \mathcal{M}_{H \to f}^{(0)*} 
\to
\texttt{M0M1}\,,
\notag \\
\mathcal{M}_{H \to f}^{(1)} \mathcal{M}_{H \to f}^{(1)*} 
&\to 
\texttt{M1M1}\,, 
\qquad
\mathcal{M}_{H \to f}^{(2)} \mathcal{M}_{H \to f}^{(0)*} 
\to
\texttt{M0M2}\,,
\end{align}
for the both $f=ggg, q\bar{q}g$.
In a similar manner, the pseudo scalar Higgs amplitudes are written as,
\begin{align}
\mathcal{M}_{q\bar{q}g}^{G,(0)} \mathcal{M}_{q\bar{q}g}^{G,(0)*}
&\to 
\texttt{MG0MG0}\,, 
\qquad
\mathcal{M}_{q\bar{q}g}^{G,(1)} \mathcal{M}_{q\bar{q}g}^{G,(0)*}
\to
\texttt{MG0MG1}\,,
\notag \\
\mathcal{M}_{q\bar{q}g}^{G,(1)} \mathcal{M}_{q\bar{q}g}^{G,(1)*}
&\to 
\texttt{MG1MG1}\,, 
\qquad
\mathcal{M}_{q\bar{q}g}^{G,(2)} \mathcal{M}_{q\bar{q}g}^{G,(0)*}
\to
\texttt{MG0MG2}\,,
\notag \\
\mathcal{M}_{q\bar{q}g}^{G,(0)} \mathcal{M}_{q\bar{q}g}^{J,(0)*}
&\to 
\texttt{MG0MJ0}\,, 
\qquad
\mathcal{M}_{q\bar{q}g}^{J,(0)} \mathcal{M}_{q\bar{q}g}^{J,(0)*}
\to
\texttt{MJ0MJ0}\,,
\notag \\
\mathcal{M}_{q\bar{q}g}^{G,(1)} \mathcal{M}_{q\bar{q}g}^{J,(0)*}
&+ \mathcal{M}_{q\bar{q}g}^{G,(0)} \mathcal{M}_{q\bar{q}g}^{J,(1)*}
\to
\texttt{MG0MJ1plusMG1MJ0}\,,
\end{align}
for $f=q\bar{q}g$. The $f=ggg$ amplitudes are written in the same fashion.
Readers can download the files and import the results into a Mathematica environment 
using \texttt{Get} or \texttt{<<}.

\bibliographystyle{JHEP}
\bibliography{AjetNNLO}

\end{document}